\documentclass{ametsocV6}

\nolinenumbers 

\title{The synchronization of convective lifecycles in an idealized microscopic model}

\authors{Hao Fu,\aff{a}\correspondingauthor{Hao Fu, haofu736@gmail.com} 
Da Yang,\aff{a} 
}

\affiliation{
\aff{a}{Department of the Geophysical Sciences, University of Chicago, Illinois}\\
}

\abstract{How a cloud ensemble responds to external forcing is a puzzle in tropical convection research. Convectively coupled gravity waves (CCGWs) in a finite domain have controllable wavelengths, providing a convenient simulation setup for studying the cloud ensemble. A multiscale analysis shows that the growth of CCGWs in a finite-domain involves not only the amplitude growth of individual clouds but also the synchronization of convective lifecycles. To understand the synchronization mechanism, we build a microscopic model with many clouds. For each cloud, the microscopic model simulates the evolution of equivalent potential temperature $\theta_e$ in the boundary layer, which is reduced by convective transport and radiative cooling and increased by surface heating. At the shallow convection stage, the $\theta_e$ grows until reaching an upper threshold where the convective inhibition energy is eliminated, and the system transitions to the deep convection stage. At the deep convection stage, the $\theta_e$ drops until reaching a lower threshold where the convective available potential energy is exhausted, and the system transitions to the shallow convection stage. The wave influences $\theta_e$ with the boundary layer convergent flow and adjusts the phase of the convective lifecycle. Numerical simulations of the microscopic model show that when the period of convection and wave equals, the wave gradually synchronizes convection. Theoretical analysis shows that the microscopic synchronization appears as the macroscopic resonant growth of the cloud ensemble. In the resonant state, the averaged $\theta_e$ and vertical velocity in the boundary layer are in phase, agreeing with the cloud-permitting simulation.  }

\begin{document}

\maketitle

%
%
%
   

  
%
%

%

\section{Introduction}

Tropical convection plays an important role in shaping the tropical climate \citep{emanuel1994book}. There are generally two perspectives to studying tropical convection. One perspective is from the microscopic level and aims to quantify the lifecycle of individual clouds \cite[e.g.,][]{byers1949thunderstorm,ogura1971numerical,ferrier1989one,morrison2020part1}. The other perspective is from the macroscopic level, which studies the statistical behavior of many clouds, i.e., a cloud ensemble \citep{arakawa1974interaction}. The cloud ensemble might be understood as a medium-size cloud population, which is large enough to yield a statistical equilibrium state and small enough to avoid background gradient in their properties.

An open question is how the cloud ensemble responds to changes in external forcing \citep{betts1986new,tompkins1998time,cohen2004response,davies2009memory}. \citet{cohen2004response} performed cloud-permitting simulations with step-changing radiative cooling and found that the cloud ensemble responds as a linear relaxation process. They revealed two adjustment timescales: a fast $O(1)$ hour timescale for gravity wave propagating a mean cloud spacing and a slow $O(10)$ days timescale for moist static energy to be vertically transported across the troposphere. \citet{davies2013departures} performed cloud-permitting simulations that periodically modulate surface heat fluxes. When the forcing period is too slow (36 hours), convection is in quasi-equilibrium with the forcing; when it is too short (1 hour), convection hardly feels the forcing; and when it is 3-12 hours, the convective amplitude varies from cycle to cycle as if there is a wave packet. \citet{davies2013departures} found that the influence of a previous convective lifecycle to the next one is carried by the lower tropospheric humidity anomalies. \citet{colin2019identifying} found that the influence is also carried by thermodynamic structures in the boundary layer, such as cold pools and convective cells. It seems necessary to include microscopic processes in understanding the response property of a cloud ensemble \cite[e.g.,][]{khouider2003coarse,colin2021atmospheric,yano2023interaction}.

Much effort has been devoted to modeling the prognostic property of a cloud ensemble, the basis for understanding its response property. \citet{pan1998cumulus} derived a linear damped oscillator model from the energetics, using convective kinetic energy, cloud-base mass flux, and cloud work function as the three governing variables. A crucial yet semi-empirical closure is a square-law relation between the convective kinetic energy and the cloud-base mass flux. \citet{yano2012ODEs} extended the closure to a general power law relation, indicating the lifecycle could obey a nonlinear oscillator. \citet{yano2012interactions} further extended the model to include both shallow and deep convection. Other studies have interpreted the dynamics of a cloud ensemble in different contexts with the broadly defined predator-prey model \citep{koren2011prey,koren2017chaos_linearized,pujol2019limit_cycle,colin2021atmospheric,yang2024predator}. It remains unclear how a cloud ensemble's lifecycle fundamentally differs from an individual cloud's, causing puzzles in designing convective parameterization schemes  \citep{emanuel1994large,mapes1997review}. Our answer is that a cloud ensemble has many clouds that might be at different lifecycle stages. This motivates us to study the \textit{phase} of a convective lifecycle and how it responds to external forcing.  



When all members of a cloud ensemble initiate and dissipate simultaneously, we call it a ``synchronized state". The only paper we know that focuses on cloud synchronization is by \citet{feingold2013coupled}. They devised a predator-prey type coupled oscillator model for mesoscale cellular convection, a shallow convective system frequently seen over subtropical oceans \citep{agee1978structure}. The governing variables are the cloud top height, droplet concentration, and rainfall rate. The neighboring clouds are assumed to couple via the downdraft-driven boundary layer flow, which is parameterized as a delayed function of the neighboring cloud's height growth rate. The cloud ensemble self-synchronizes when the coupling strength parameter takes a moderate value. This model impressively demonstrates individual convection's phase, period, and amplitude adjustment. It carries many regimes that await further study. However, it is probably too complicated to yield an analytical solution, preventing further interpretation of physics. The physical basis of this model involves some empirical considerations to fit it into the classic predator-prey type model. Whether this model can be applied to tropical convection also remains unclear. Thus, we decided to devise a simpler model focusing on the convective lifecycle phase. It should analytically reveal the statistical behavior of synchronization and how synchronization could bridge the microscopic and macroscopic cloud dynamics. We also need an objective index to quantify synchronization in observational data, cloud-permitting simulations, or simple models.

This paper is organized in the following way. In section \ref{sec:RCE}, we introduce an example problem, the convectively coupled gravity waves (CCGWs) simulated in a finite domain, for understanding the response of a cloud ensemble to periodic external forcing. Multiscale analysis shows that the growth of CCGW is significantly contributed by the synchronization of convective lifecycles, which can be quantified with an index. In section \ref{sec:micro_model}, we propose a simple microscopic model of the convective lifecycle to reproduce the synchronization behavior. The microscopic model has a fixed amplitude, but the wave-induced large-scale vertical motion can adjust its phase. An analytical solution of the model shows that the phase adjustment of individual clouds makes the cloud ensemble an approximate \textit{harmonic oscillator}. Section \ref{sec:discussion} discusses the limitations of the simple model, followed by a conclusion in section \ref{sec:conclusion}.

\section{Evidence of synchronization in an RCE simulation
}\label{sec:RCE}

\subsection{Experimental setup}\label{subsec:RCE_exp}

We simulate CCGW in an idealized configuration, the radiative-convective equilibrium (RCE) setup. It is a simulation configuration with no Coriolis force, no initial background wind, a uniform sea surface temperature, and periodic boundary conditions in the two horizontal directions. The RCE setup is widely used in studying the statistical behavior of tropical convection \cite[e.g.,][]{held1993radiative,tompkins2001coldpool,bretherton2005energy,romps2014analytical,wing2018convective}. CCGWs have been reported in many RCE simulations, with or without Coriolis force \citep{bretherton2005energy,bretherton2006interpretation,nolan2007trigger,tulich2007vertical,o2017accessible,yang2019convective,yang2020interactive,fu2023small}. The waves are usually in a standing pattern caused by the interference of counter propagating waves. Their wavelength is severely constrained by the domain size, if not wide enough. As a result, the standing wave has been attributed to be largely irrelevant to the real atmosphere. We argue that this limitation can be useful for specific problems, such as studying the response of a cloud ensemble to waves with controllable wavelength and frequency. We will report the convective synchronization phenomenon in a finite-domain RCE simulation, with an alert that the wave could be different from CCGWs in the real atmosphere. 


We use Cloud Model 1 \cite[CM1, version 19.10,][]{bryan2002benchmark} to perform a cloud-permitting RCE simulation in a 1080 km by 1080 km doubly periodic square domain with a fixed sea surface temperature of 300 K. The setup is identical to that used by \citet{fu2023small} except for setting zero Coriolis parameter here. The domain height is 27 km, with a sponge layer above 20 km to attenuate gravity waves. The mesh is vertically nonuniform, with eight mesh layers in the lowest 1 km to resolve the boundary layer processes. The physics parameterization package includes the Morrison double-moment microphysics scheme \citep{morrison2005new}, the RRTMG radiative transfer scheme \citep{clough2005RRTM}, the surface model based on the similarity theory \citep{jimenez2012revised}, and the simple boundary layer scheme from \citet{bryan2009maximum}. To remove the influence of the diurnal cycle, the solar zenith angle is set a constant. The simulation is initiated with an RCE-state sounding plus random noise of potential temperature in the lowest five mesh layers and runs for 10 days. The data output frequency is every 0.5 hours. The RCE-state sounding is the last day output of a 100-day small-domain simulation with a horizontal domain size of 120 km $\times$ 120 km. Next, we analyze the simulation results from a macroscopic and microscopic perspective and try reconciling them.

\subsection{Macroscopic analysis}
\label{subsec:macro}

\begin{figure}[h]
\centerline{\includegraphics[width=1\linewidth]{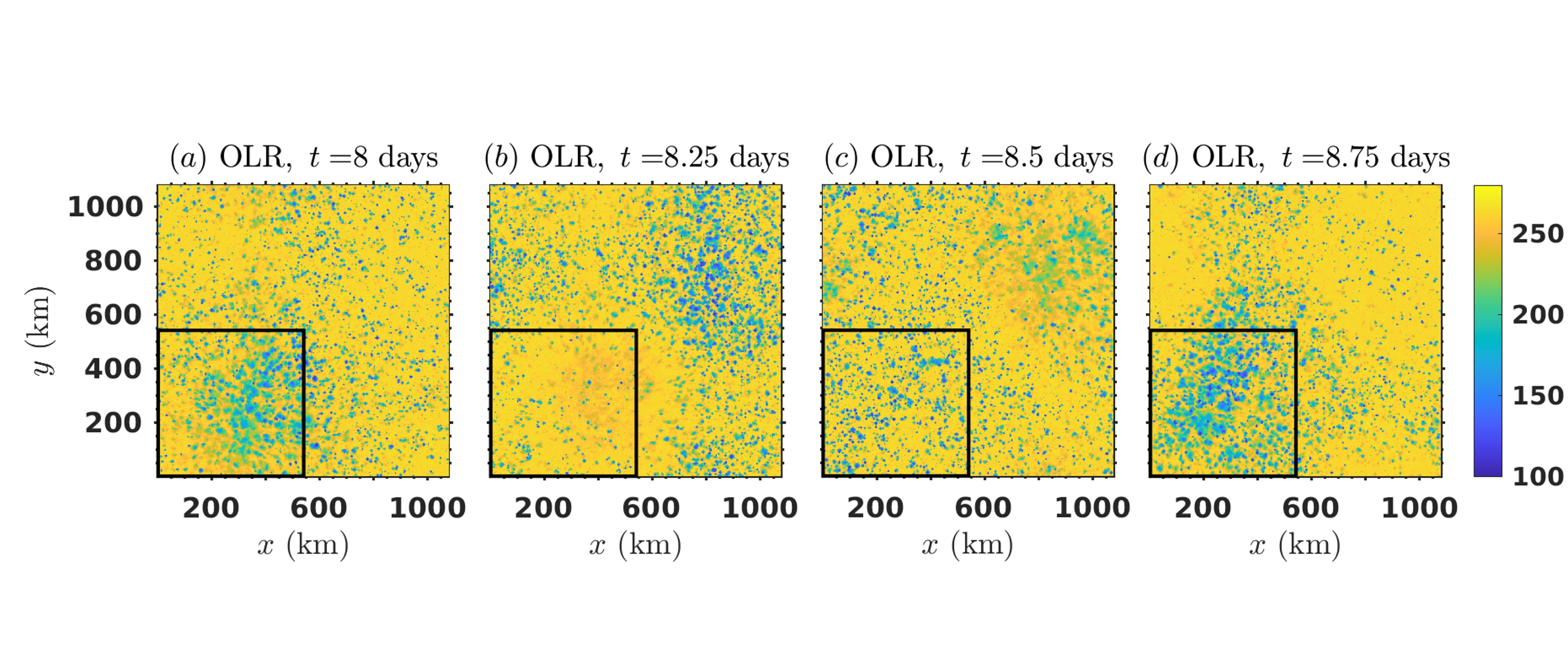}}
\caption{The outgoing longwave radiation (unit: W m$^{-2}$) of the cloud-permitting simulation at (a) $t=8$ days, (b) $t=8.25$ days, (c) $t=8.5$ days, and (d) $t=8.75$ days. They roughly cover one period of the standing wave. The OLR field has been shifted to locate the most oscillatory block at the southwest corner, as outlined by the black box. The boxed region has a size of 540 km $\times$ 540 km. A movie version is in the supplemental material. }   \label{fig:2Dxy_OLR}
\end{figure}

At the macroscopic level, we analyze the wave's vertical structure and phase relation between different quantities. Because the standing CCGWs in the RCE setup have not been sufficiently investigated, particular attention is paid to how they align with or differ from the CCGWs in the literature. We analyze the domain-scale wave with a horizontal total wavenumber of $\sqrt{2} (2\pi/L)$, which has a four-quadrants pattern (Fig. \ref{fig:2Dxy_OLR}). Here, $L=1080$ km is the domain width. The domain is split into four blocks. To avoid coincidentally selecting the wave node with nearly zero amplitude, we have rearranged the figure such that the most oscillatory regions are along the diagonals. Most variables analyzed in this subsection are horizontal averages within the boxed region, denoted with a ``$\widetilde{\quad}$" operator. It is essentially the average over a half wavelength of the standing wave, across which the wave phase is relatively uniform. The cloud population inside can be viewed as a cloud ensemble.



\begin{figure}[h]
 \centerline{\includegraphics[width=1\linewidth]{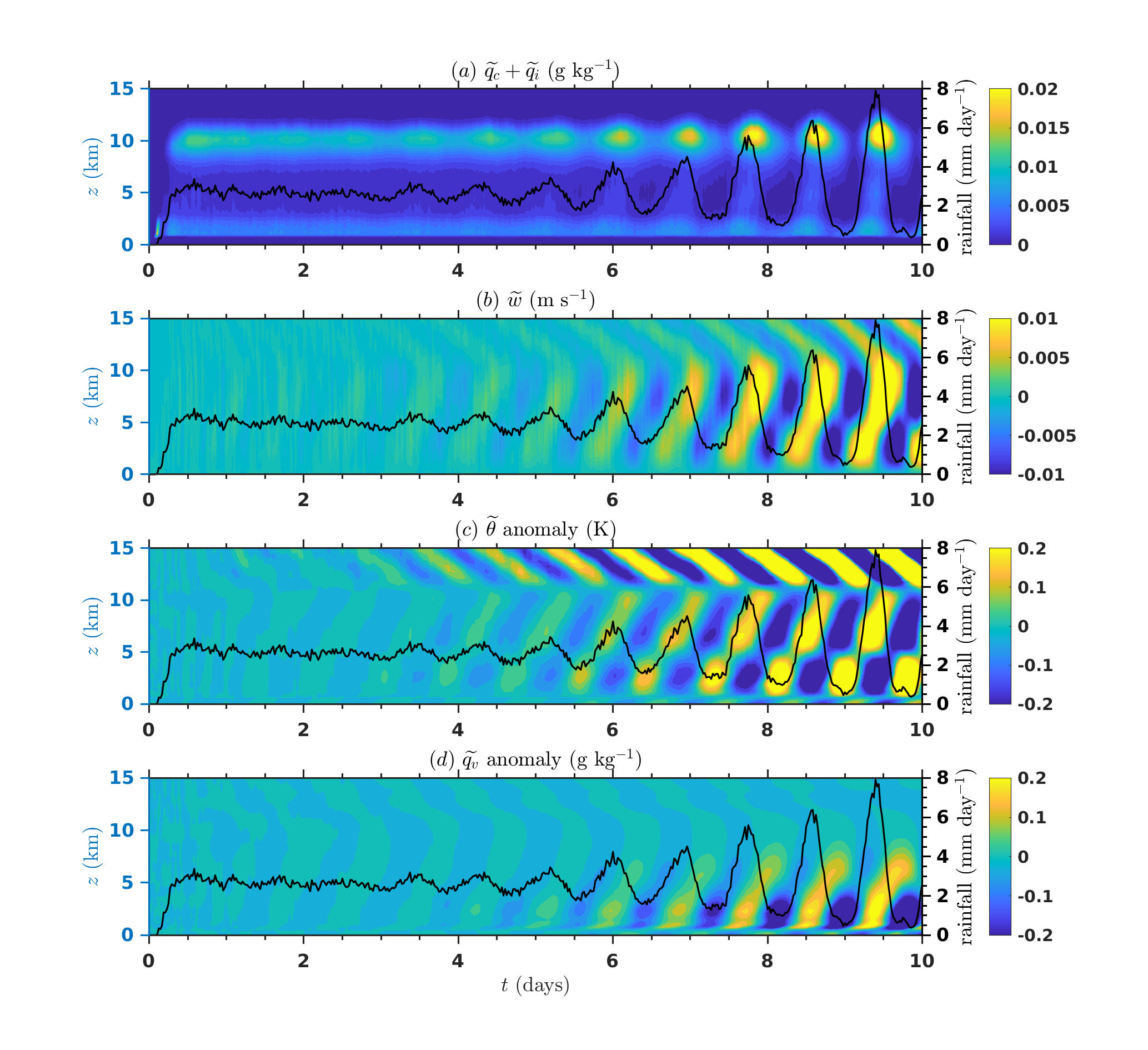}}
\caption{The time-height diagram of box-averaged quantities in the cloud-permitting simulation. (a) The sum of the cloud liquid water and ice mixing ratio. (b) The vertical velocity. (c) The potential temperature subtracting the domain horizontal average. (d) The water vapor mixing ratio subtracting the domain horizontal average. The black lines show the time series of the box-averaged surface rainfall rate.}   \label{fig:2Dxz_all}
\end{figure}

Figure \ref{fig:2Dxz_all} shows the time evolution of the wave's vertical structure. All quantities are box-averaged values. The sum of the cloud liquid and ice water mixing ratio $\widetilde{q_c}+\widetilde{q_i}$ (Fig. \ref{fig:2Dxz_all}a) shows the clouds mainly have two types: 1) shallow cumulus and congestus clouds between $z=1$ km and 3 km, and 2) stratiform clouds between $z=9$ km and 12 km. The shallow clouds lead the stratiform clouds. There is a transition zone between the two cloud types (e.g., $t=7.75$ days), which we infer to be deep convective clouds. The precipitation peaks at the deep convective stage. This structure agrees with the previous understanding of the lifecycle of a mesoscale convective system \cite[e.g.,][]{kiladis2009convectively}. The following analysis will use the surface rainfall rate as a reference for the wave phase. The main features are summarized below:
\begin{itemize}
    \item Figure \ref{fig:2Dxz_all}b shows the lower-tropospheric vertical velocity leads the surface rainfall rate by $\pi/2$ phase and is in phase with shallow convection. Shallow convection may respond to boundary layer convergence almost instantaneously and be the preconditioner for deep convection, thereby leading precipitation. \cite[e.g.,][]{khouider2006simple,liu2022convective}. 
    \item Figure \ref{fig:2Dxz_all}c shows that the horizontal anomaly of lower-tropospheric potential temperature $\widetilde{\theta}$ is in the opposite phase of the rainfall. The second baroclinic mode dominates the vertical structure of $\widetilde{\theta}$ below 12 km height.\footnote{The vertical structure of $\widetilde{\theta}$ is different from $\widetilde{w}$. The latter manifests both the first and second baroclinic modes. The reason is that the first-mode wave has a faster intrinsic wave speed that spreads its temperature anomaly more smoothly in the domain than the second mode \citep{wu2003shallow}.} Heavy precipitation, both due to deep convection and stratiform clouds, is in phase with a low-level cold anomaly, which reduces the low-level static stability and could invigorate convection. This indicates the presence of stratiform instability \citep{mapes2000toy,majda2001models,bretherton2005energy}.
    \item Figure \ref{fig:2Dxz_all}d shows the water vapor mixing ratio $\widetilde{q_v}$ has a distinct anomaly at the BL top (around 1 km high), which is in phase with shallow convection, and another anomaly at around 3 km, which is in phase with deep convection. This agrees with the previous understanding that the transition from shallow to deep convection is accompanied by the gradual moistening of the lower troposphere \citep{khouider2006simple}. 
\end{itemize}

 
The above analysis shows that the standing wave in the RCE simulation is consistent with previous understandings of CCGW. This paper does not aim to model the wave growth rate. Instead, we use the CCGW as a lens to study the response of a cloud ensemble to large-scale vertical motion.

\begin{figure}[h] \centerline{\includegraphics[width=1.2\linewidth]{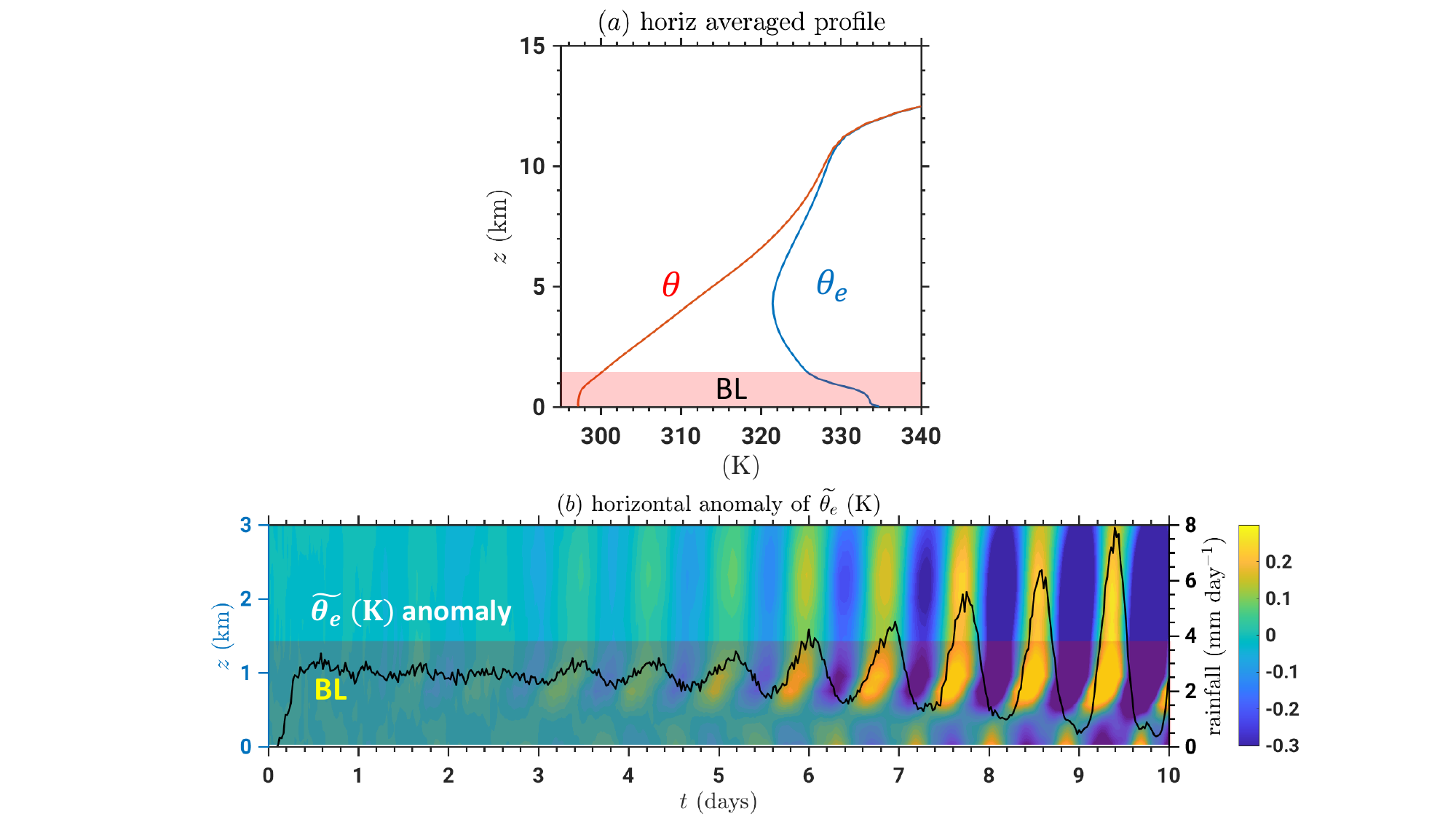}}
\caption{(a) The blue line shows the vertical profile of the domain horizontally averaged $\theta_e$ at $t=4$ days. The red shading region denotes the boundary layer region, defined as between $z=0$ km and $z=H_B=1.43$ km. (b) The time-height diagram of the mean equivalent potential temperature ($\widetilde{\theta_e}$) of the boxed region, subtracting the domain horizontal average. The diagram zooms into the lowest 3 km. The black line shows the time series of the box-averaged surface rainfall rate. The shaded region denotes the boundary layer.
}   \label{fig:the_profile}
\end{figure}

The equivalent potential temperature $\theta_e$ is a bulk measure of humidity and buoyancy, a quantity closely related to the moist static energy (MSE) in the lower troposphere \citep{riehl1979climate}. The $\theta_e$ is a conserved variable in water phase change and a convenient tool for analyzing the convective lifecycle. Without considering the ice phase process, $\theta_e$ can be approximated as \cite[e.g.,][]{siebesma1998shallow}:
\begin{equation}  \label{eq:thetae_simplified}
    \theta_e \approx \theta \exp \left( \frac{L_v q_v}{c_p T} \right),
\end{equation}
where $T$ is temperature, $L_v \approx 2.5\times10^6$ J kg$^{-1}$ is the specific latent heat of water, and $c_p \approx 1005$ J kg$^{-1}$ K$^{-1}$ is the isobaric specific heat of air. Traditionally, the $\theta_e$ or MSE in the mixed layer, i.e., lowest 0.5 km, has been used to represent the potential for convective initiation \citep{mapes1993gregarious,raymond1995BLQE,colin2021atmospheric}. However, \citet{fuglestvedt2020cold} showed that moistening of the upper boundary layer (0.8 km to 1 km height) is also crucial for deep convective initiation. In our simulation, both $\widetilde{q_v}$ and $\widetilde{\theta_e}$ have a distinct anomaly between 0.5 km and 1.4 km height (Fig. \ref{fig:the_profile}b), which overlaps with the region of high vertical gradient of the background $\theta_e$ (Fig. \ref{fig:the_profile}a). This region is likely the entrainment zone, a transition layer between the mixed layer and the free troposphere \cite[e.g.,][]{stull2012introduction}. Thus, we focus on the $\theta_e$ anomaly averaged between $z=0$ km and $z=H_B=1.43$ km, the lowest ten mesh layers. It is the boundary layer in a generalized sense, which includes the mixed layer and the entrainment zone. We hypothesize that the vertically averaged $\theta_e$ in the boundary layer is a crucial indicator of convective initiation and termination. Next, we use a microscopic analysis to examine the distribution of $\theta_e$ within the boxed region. We have performed microscopic analyses with two other choices of $H_B$ with a change of around -20\% and +40\%: $H_B=1.18$ km and $H_B=2.01$ km. They do not exhibit qualitative differences. Thus, we will only report the result calculated with $H_B=1.43$ km.

\subsection{Microscopic analysis}
\label{subsec:micro}

The boxed region is a cloud ensemble that contains many clouds. We use a coarse-graining technique to analyze the convective lifecycle at the microscopic level. The data in the boxed region is uniformly divided into 20 km $\times$ 20 km cells, with each cell representing a convective cloud and its associated cold pool. For the $n^{th}$ convective cell, the horizontal velocity ($u_n$, $v_n$), vertical velocity ($w_n$), and equivalent potential temperature ($\theta_{e,n}$) are decomposed into:
\begin{equation}  \label{eq:symbol_decomposition}
\begin{split}
    u_n(x,y,z,t) &= \overline{u_n}(z,t) + u'_n (x,y,z,t), \\
    v_n(x,y,z,t) &= \overline{v_n}(z,t) + v'_n (x,y,z,t), \\ 
    w_n(x,y,z,t) &= \overline{w_n}(z,t) + w'_n (x,y,z,t), \\   
    \theta_{e,n}(x,y,z,t) &= \overline{\theta_{e,n}}(z,t) + \theta'_{e,n} (x,y,z,t), \\       
\end{split}    
\end{equation}
where $x$ and $y$ denote the variation within a convective cell. An overbar denotes performing horizontal average within a cell:
\begin{equation}
    \overline{\theta_{e,n}} \equiv \frac{1}{(\Delta x)^2} \iint \theta_e dx dy,
\end{equation}
and a variable with a prime denotes the anomaly with respect to the cell average. Here $\Delta x=20$ km is the cell width. The $\overline{\theta_{e,n}}$ is further decomposed into a background part $\theta_{e,0}$, taken as the domain horizontal average of $\theta_e$, and an anomalous part. The anomalous part in the BL is defined as $\Theta_n$, calculated as the vertical average between $z=0$ km and $z=H_B=1.43$ km. Similarly, we define $W_n$ as the mean $\overline{w_n}$ in the BL:
\begin{equation}
\begin{split}
    \Theta_n &\equiv \langle \overline{\theta_{e,n}} \rangle - \langle \theta_{e,0} \rangle, \\
    W_n &\equiv \langle \overline{w_n} \rangle,
\end{split}    
\end{equation}
where $\langle \rangle$ denotes vertical average in the boundary layer:
\begin{equation}
    \langle \rangle \equiv \frac{1}{H_B} \int_{0}^{H_B} dz.
\end{equation}

\begin{figure}[h] \centerline{\includegraphics[width=1.2\linewidth]{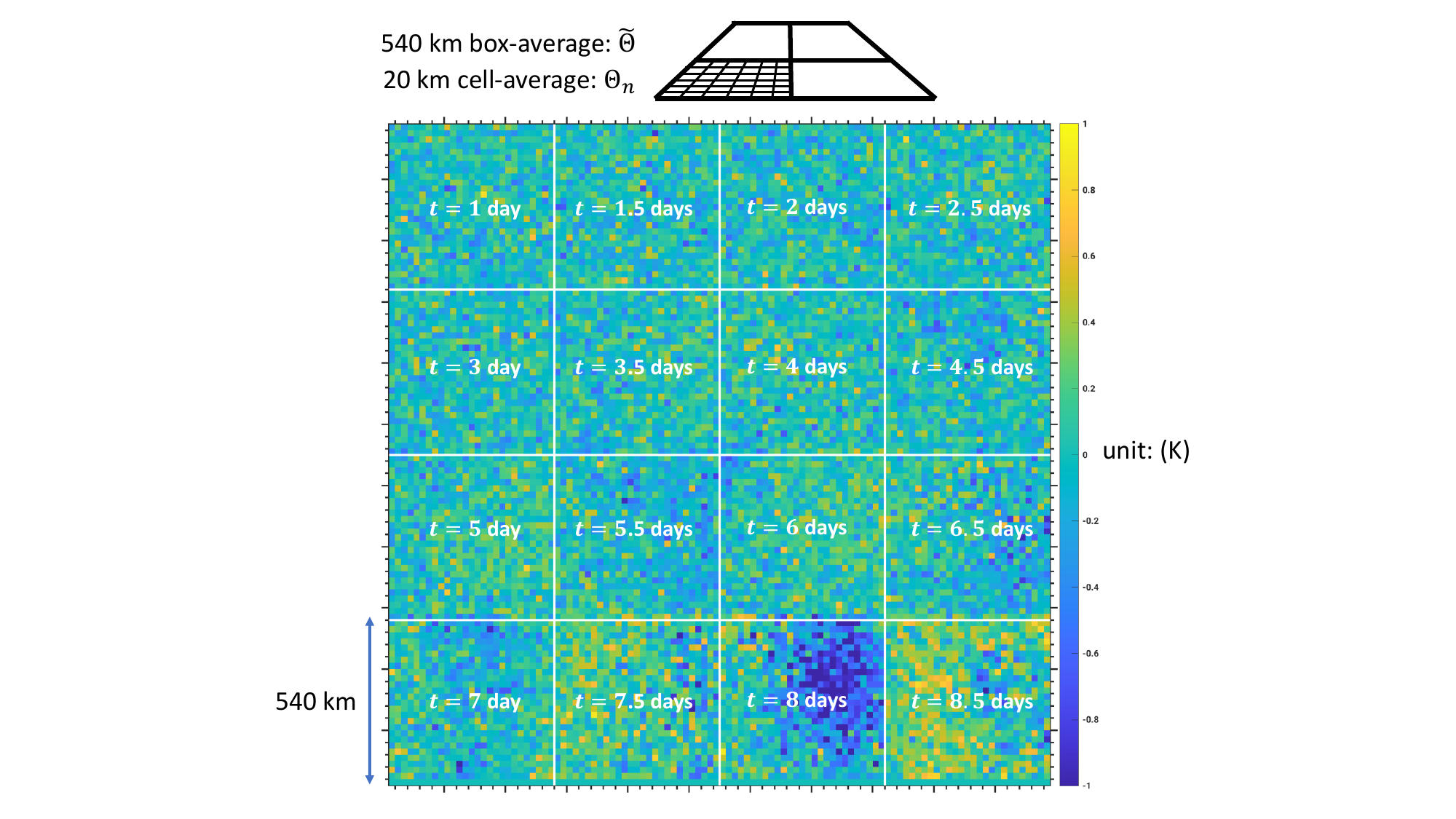}}
\caption{The $\Theta_n$ at 16 snapshots sampled between $t=1$ day and $t=8.5$ days of the cloud-permitting simulation. It is the boundary layer-averaged $\theta_e$ that has subtracted the domain horizontal average and has been coarse-grained. Each pixel represents the spatial average within a 20 km $\times$ 20 km cell in the boxed region (540 km $\times$ 540 km). The 540 km is half the domain width ($L=1080$ km). A movie version is in the supplemental material. }   \label{fig:checkerboard}
\end{figure}

\begin{figure}[h]
 \centerline{\includegraphics[width=0.7\linewidth]{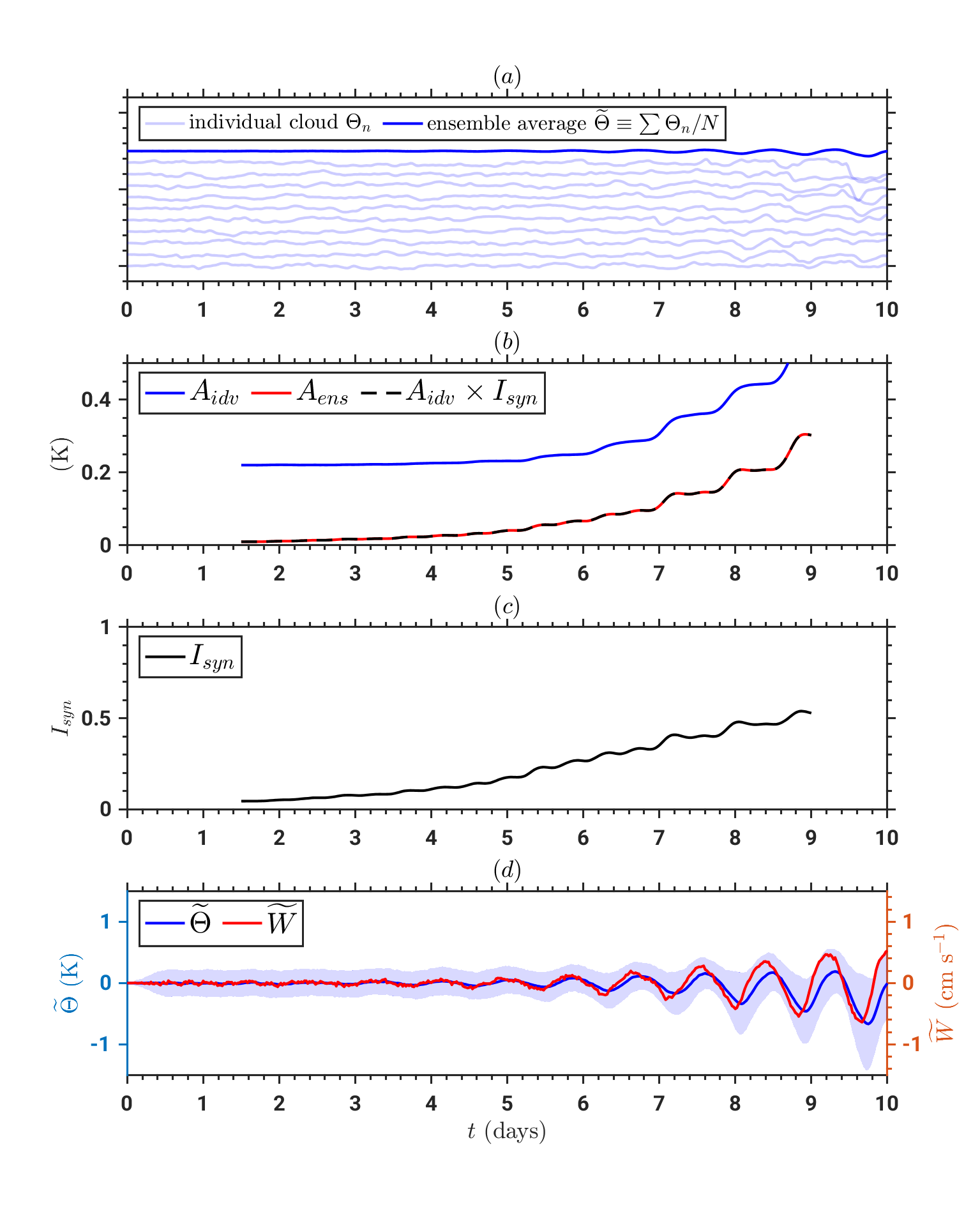}}
\caption{(a) The shallow blue lines show the BL-averaged equivalent potential temperature perturbation ($\Theta_n$) of ten randomly sampled coarse-grained cells. To facilitate visualization, each time series has been smoothed by a 3-hour-scale Gaussian filter. The dark blue line shows the ensemble average of all cells in the boxed region, $\widetilde{\Theta}$. The y-axis scale is not shown. (b) The time series of the individual cell amplitude $A_{idv}$ (solid blue line), the cloud ensemble amplitude $A_{ens}$ (solid red line), and the product of $A_{idv}$ and $I_{syn}$ (dashed black line). (c) The time series of the synchronization index $I_{syn}$. (d) The blue line shows $\widetilde{\Theta}$, and the red line shows the ensemble-averaged-and-BL-averaged vertical velocity $\widetilde{W}$ (in the unit of cm s$^{-1}$). The blue shading shows the $\pm1$ standard deviation range of the cloud ensemble's $\Theta_n$, i.e., the spatial standard deviation over the coarse-grained data. }   \label{fig:synchro_time_series}
\end{figure}   


Figure \ref{fig:checkerboard} shows that $\Theta_n$ takes a mosaic pattern within the first few days, fluctuating by about 1 K. The CCGW manifests as an emerging synchronized pattern, as well as a higher oscillatory amplitude in each cell. Figure \ref{fig:synchro_time_series}a displays the time series of $\Theta_n$ from 10 randomly sampled cells, confirming the co-existence of amplitude growth and synchronization, which both contribute to the growth of the ensemble (box-averaged) quantity $\widetilde{\Theta}$. We introduce a method to decompose the cloud ensemble amplitude into the amplitude of individual cells and the synchronization effect. We let 
\begin{equation}
\Theta_{np},\,\,n=1,\,\,2,\,\, ..., N; \,\, p = 1, \,\,2, ..., \,\,P,  
\end{equation}
be a 2-D array of cell-averaged equivalent potential temperature, where $N=(540\,\,\mathrm{km}/20\,\,\mathrm{km})^2=729$ is the ensemble size (Fig. \ref{fig:checkerboard}), and $P$ is the number of time snapshots for the calculation. Because a wave period is around a day, we use a data length of 2 days to capture the wave signal, which includes $P=97$ time snapshots. To avoid sampling the initial spin-up stage of the simulation, the center of the calculation window starts at $t=1.5$ days and ends at $t=9$ days. The cloud ensemble amplitude, $A_{ens}$, is calculated by the temporal square average of $\widetilde{\Theta}$ in the calculation slot:
\begin{equation} \label{eq:A_wave}
    A_{ens} \equiv 
    \sqrt{ \frac{1}{P} \sum_p \left( \frac{1}{N} \sum_n \Theta_{np} \right)^2 },
\end{equation}
where $\widetilde{\Theta} \equiv \frac{1}{N} \sum_n \Theta_{np} $. The mean amplitude of individual cell, $A_{idv}$, is obtained by calculating the temporal square average of individual cloud first, and then averaging over the cloud ensemble:
\begin{equation} 
\label{eq:A_idv}
\begin{split}
    A_{idv}
    \equiv \sqrt{ \frac{1}{NP} \sum_n \sum_p \Theta^2_{np} }.
\end{split}    
\end{equation}
Figure \ref{fig:synchro_time_series}b shows that $A_{idv}$ amplifies by 100\% within nine days. Finally, we introduce the synchronization index $I_{syn}$, which uses the spatial variance normalized by $A^2_{idv}$:
\begin{equation} \label{eq:Isyn}
\begin{split}
    I_{syn}
    &\equiv \left( 1 - \frac{\text{spatial var}}{A^2_{idv}} \right)^{1/2} \\
    &=\left[ 1 - \frac{ \frac{1}{NP} \sum_p \sum_n \left( \Theta_{np} - \frac{1}{N}\sum_{n'} \Theta_{n'p} \right)^2 }{ \frac{1}{NP} \sum_p \sum_n \Theta_{np}^2 } \right]^{1/2}.
\end{split}    
\end{equation}
The $I_{syn}$ ranges between 0 and 1, with $I_{syn}=0$ representing a fully incoherent state with random phase between clouds, and $I_{syn}=1$ representing a fully synchronized state. Figure \ref{fig:synchro_time_series}c shows that $I_{syn}$ increases from around zero to 0.5. In Appendix A, we show:
\begin{equation}  \label{eq:wave_decomposition}
    A_{ens} = A_{idv} \times I_{syn}.
\end{equation}
Applying this assumption to the definition of $A_{ens}$ and $A_{idv}$,  Figure \ref{fig:synchro_time_series}b validates Eq. (\ref{eq:wave_decomposition}). The diagnostic result shows that the growth of $A_{idv}$ and $I_{syn}$ cooperatively contribute to the growth of CCGW. However, synchronization is probably more fundamental because $I_{syn}$ grows from nearly zero.



Figure \ref{fig:synchro_time_series}d shows that the ensemble-averaged vertical velocity in the BL, $\widetilde{W}$, is almost in phase with $\widetilde{\Theta}$ within the ten-day-long simulation period.\footnote{After $t=7$ days, $\widetilde{W}$ begins to slightly lead $\widetilde{\Theta}$, a phenomenon we cannot explain now. } This property will be reproduced by the microscopic model in section \ref{sec:micro_model} and interpreted as a resonance between the wave and the cloud ensemble.

\section{Microscopic model: a dual-threshold system}\label{sec:micro_model}

In the last section, we found that the growth of CCGWs in the finite-domain simulation is significantly contributed by the synchronization of convective lifecycles, and quantified it with a synchronization index. In this section, we build a simple model to understand the mechanism of convective synchronization under a prescribed wave forcing. The model setup and simulation results are introduced in sections \ref{sec:micro_model}\ref{subsec:derivation_micro} and \ref{sec:micro_model}\ref{subsec:microscopic_result}, and an approximate analytical solution is presented in section \ref{sec:micro_model}\ref{subsec:oscillator}.

\subsection{Derivation of the microscopic model}\label{subsec:derivation_micro}

As explained in section \ref{sec:RCE}\ref{subsec:macro}, we use the vertically averaged $\theta_e$ in the BL as the control variable of the microscopic convective lifecycle model. The first step is to derive the governing equation of $\Theta_n$, representing the horizontal anomaly of the BL-averaged $\theta_e$. Using the conservation law of $\theta_e$, we obtain:
\begin{equation}  \label{eq:Theta_raw}
    \frac{\partial \Theta_n}{\partial t}
    = - \left\langle \overline{u_n \frac{\partial \theta_{e,n}}{\partial x} } \right\rangle
      - \left\langle \overline{v_n \frac{\partial \theta_{e,n}}{\partial y}} \right\rangle
      - \left\langle \overline{w_n \frac{\partial \theta_{e,n}}{\partial z}} \right\rangle
      + \left\langle \overline{Q_{rad}} \right\rangle
      - \left\langle \overline{ \frac{\partial F_{sgs}}{\partial z}} \right\rangle - \frac{d\langle \theta_{e,0}\rangle}{dt},
\end{equation}
where $Q_{rad}$ is the radiative heating rate, $d\langle \theta_{e,0} \rangle / dt$ is the tendency of the domain-averaged $\theta_e$ in the BL, and $F_{sgs}$ represents the subgrid vertical fluxes in the BL. Because the $z=H_B=1.43$ km is well above the mixed layer, subgrid fluxes almost vanish at this height. Thus, the vertical average of $F_{sgs}$ obeys:
\begin{equation}
- \left\langle \overline{ \frac{\partial F_{sgs}}{\partial z}} \right\rangle
    \approx \overline{Q_{surf}},
\end{equation}
where $Q_{surf}$ is the surface flux of $\theta_e$ divided by $H_B$. We approximate the surface flux of $\theta_e$ as the surface flux of $\theta + (L_v/c_p)q_v$ divided by the boundary layer depth $H_B$. Here, $\theta + (L_v/c_p)q_v$ is the linearized expression of $\theta_e$ at the surface.\footnote{Using $L_v=2.5\times10^6$ J g$^{-1}$, $c_p=1005$ J kg$^{-1}$ K$^{-1}$, a characteristic $q_v$ of 0.015 kg kg$^{-1}$ and $T$ of 300 K, we get $L_v q_v/(c_p T) \approx 0.12$, so the linearization is valid. }

Next, we perform a decomposition of Eq. (\ref{eq:Theta_raw}) into 20 km $\times$ 20 km cell-averaged quantities and perturbation quantities, with each cell representing a cloud:
\begin{equation}  \label{eq:Theta_decompose}
\begin{split}
    \frac{\partial \Theta_n}{\partial t}
    &= - \left\langle \overline{u_n} \overline{\frac{\partial \theta_{e,n}}{\partial x} } \right\rangle
    - \left\langle \overline{v_n} \overline{\frac{\partial \theta_{e,n}}{\partial y} } \right\rangle
    - \left\langle \overline{w_n} \overline{\frac{\partial \theta_{e,n}}{\partial z} } \right\rangle \\
    &\quad - \left\langle \overline{u'_n {\frac{\partial \theta'_{e,n}}{\partial x} }} \right\rangle
    - \left\langle \overline{v'_n {\frac{\partial \theta'_{e,n}}{\partial y} }} \right\rangle
    - \left\langle \overline{w'_n {\frac{\partial \theta'_{e,n}}{\partial z} }} \right\rangle \\
    &\quad + \left\langle \overline{Q_{rad}} \right\rangle
    +  \overline{Q_{surf}} 
    - \frac{d\langle \theta_{e,0}\rangle}{dt}.
\end{split}      
\end{equation}
Because the magnitude of $\Theta_n$ is on the order of 1 K (Fig. \ref{fig:checkerboard}), much smaller than the roughly 9 K lapse of $\theta_e$ across the boundary layer (Fig. \ref{fig:the_profile}a), the vertical advection of the background $\theta_e$ gradient should dominate the vertical advection term:
\begin{equation}  \label{eq:assumptions_Theta}
\begin{split}
    O\left( \overline{w_n} \frac{\partial \overline{\theta_{e,n}} }{\partial z} \right) 
    \approx 
    O \left( \overline{w_n} \frac{d\theta_{e,0}}{dz} \right). 
\end{split}
\end{equation}
Budget analysis in Appendix B suggests:
\begin{equation}  \label{eq:assumptions_Theta_horizontal}
\begin{split}
    O\left( \overline{u_n} \overline{\frac{\partial \theta_{e,n}}{\partial x}} \right),
    \,\,
    O\left( \overline{v_n} \overline{\frac{\partial \theta_{e,n}}{\partial y}} \right)
    \,\,    
    \ll 
    \,\, O \left( \overline{w_n} \frac{d\theta_{e,0}}{dz} \right), 
\end{split}
\end{equation}
as well as the approximation of $Q_{rad}$ as a negative constant and $Q_{surf}$ as a positive constant. The budget also shows that the background tendency term $-d\langle \theta_{e,0}\rangle/dt$ can be neglected within the first 10 days. 


Substituting Eqs. (\ref{eq:assumptions_Theta}) and (\ref{eq:assumptions_Theta_horizontal}) into Eq. (\ref{eq:Theta_raw}), and dropping the background tendency term, we get:
\begin{equation}  \label{eq:Theta_simplified}
    \frac{d \Theta_n}{dt}
    \approx \underbrace{ - \widetilde{W} \left\langle \frac{d\theta_{e,0}}{dz} \right\rangle }_{\text{wave}}\,\,\, 
    \underbrace{ 
    - \left\langle \overline{u'_n \frac{\partial \theta'_{e,n}}{\partial x}}
    + \overline{v'_n \frac{\partial \theta'_{e,n}}{\partial y}}
    + \overline{w'_n \frac{\partial \theta'_{e,n}}{\partial z}} 
    + \left( \overline{w_n} - \widetilde{W} \right) \frac{d\theta_{e,0}}{dz}  
    \right\rangle 
    }_{\text{convection}} 
    \,\,\, + \,\,\, \underbrace{Q_{rad}}_{\text{radiation}}
    \,\,\, + \,\,\, \underbrace{Q_{surf}}_{\text{surf heating}}.    
\end{equation}
Equation (\ref{eq:Theta_simplified}) has a wave forcing term, a convection term, a radiative cooling term, and a surface heating term. We remark that the split into wave and convection terms is not rigorous. The gravity wave is excited by diabatic heating, while convection is also sustained by diabatic heating. Thus, the cell-averaged vertical velocity, $\overline{w_n}$, is contributed by both wave and convection. Because the domain-scale CCGW has a much wider horizontal scale than an individual cloud, we perform an approximate split by letting $\widetilde{W}$ represent the wave vertical motion and letting $\overline{w_n}-\widetilde{W}$ represent the convective vertical motion.

\begin{figure}[h]
\centerline{\includegraphics[width=1\linewidth]{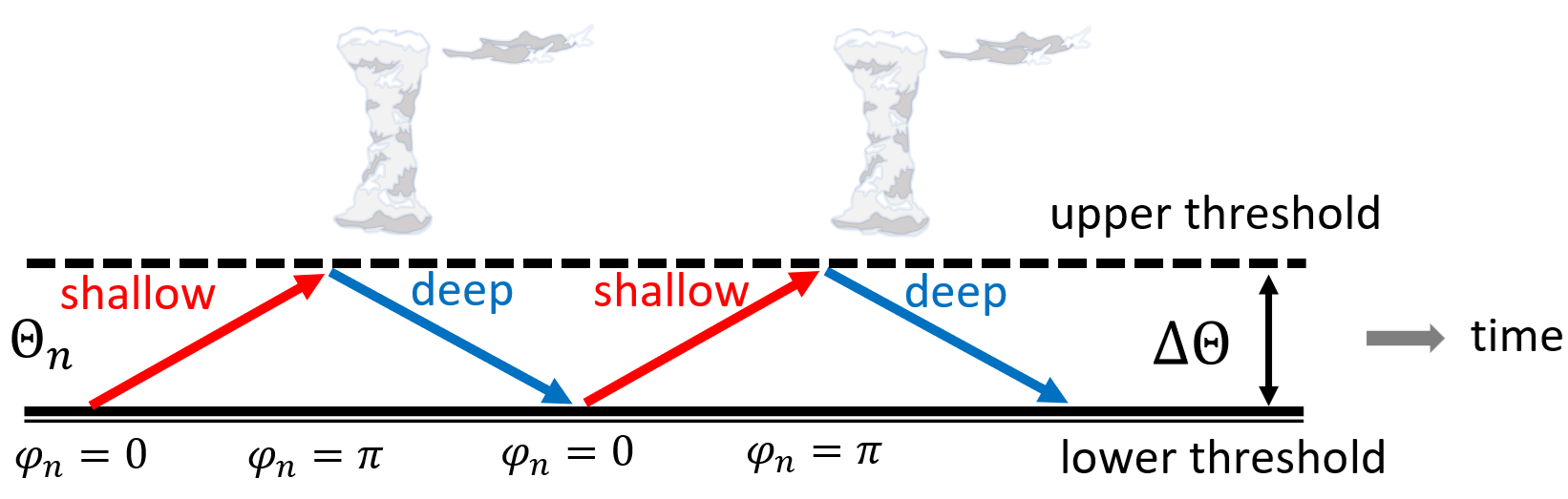}}
\caption{A schematic diagram of the microscopic model. The convective lifecycle is controlled by the anomalous equivalent potential temperature in the BL, $\Theta_n$, with $n=$1, 2, 3, ... denoting the $n^{th}$ cell. The amplitude of $\Theta_n$ is set by the lower and an upper threshold, whose interval is $\Delta \Theta$. The shallow stage is characterized by shallow cumuli and congestus convection; the deep stage is characterized by deep convection and stratiform clouds. The deep-to-shallow transition point is defined as the $\varphi_n=0$ phase, and the shallow-to-deep transition point is defined as the $\varphi_n=\pi$ phase. The wave forcing influences the phase rather than the amplitude of an individual cloud.}   \label{fig:cartoon_dual_threshold}
\end{figure}

\begin{figure}[h]
\centerline{\includegraphics[width=1.05\linewidth]{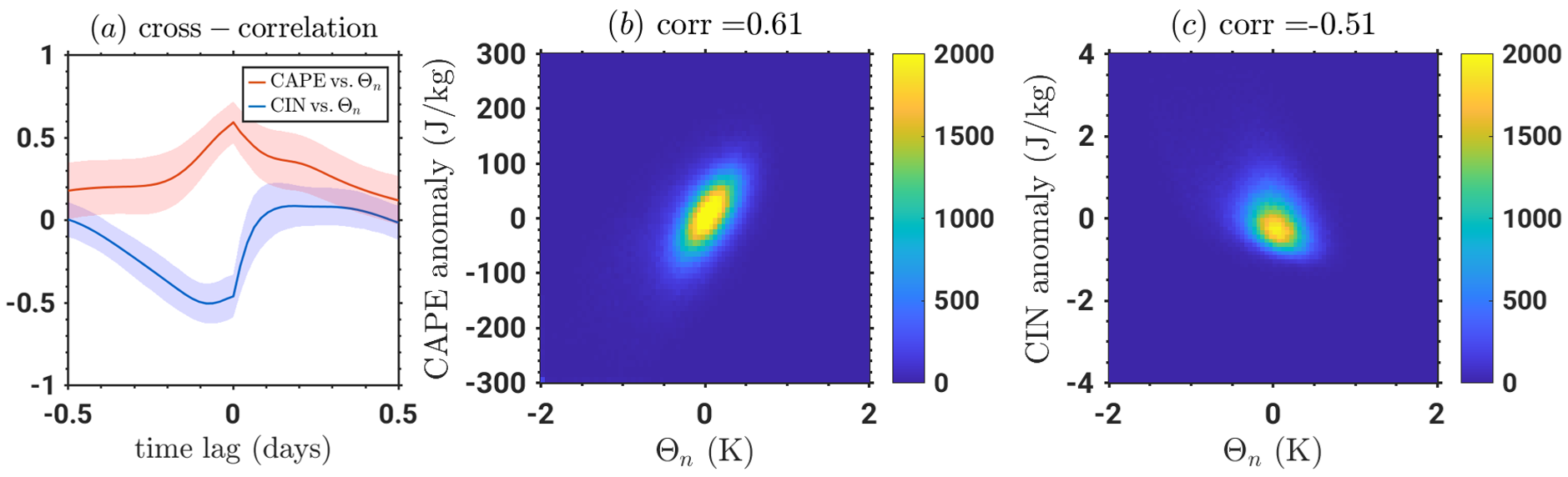}}
\caption{(a) The solid red line shows the ensemble-averaged cross-correlation between CAPE and $\Theta_n$, using different time lags. The solid blue line shows the ensemble-averaged cross-correlation between CIN and $\Theta_n$. Both CAPE and CIN are anomalies that have subtracted the domain-averaged values. The shading shows the $\pm1$ standard deviation of the cross-correlation over the ensemble members. The time series between $t=1$ day and $t=10$ days are used, with an interval of 0.5 hours. The time lag is defined so that if the cross-correlation profile peaked at a negative time lag, it would mean CAPE or CIN leads $\Theta_n$. (b) and (c) show the density distribution of $\Theta_n$ in space and time versus the anomalous CAPE and CIN, respectively. The colorbar shows the number of data points in each pixel. The correlation coefficients are shown in the titles. } \label{fig:lag}
\end{figure}

The convection term is decomposed into the shallow and deep parts, $Q_{s,n}$ and $Q_{d,n}$:
\begin{equation}  \label{eq:Fsn_Fdn}
  - \left\langle \overline{u'_n \frac{\partial \theta'_{e,n}}{\partial x}}
    + \overline{v'_n \frac{\partial \theta'_{e,n}}{\partial y}}
    + \overline{w'_n \frac{\partial \theta'_{e,n}}{\partial z}}
    + \left( \overline{w_n} - \widetilde{W} \right) \frac{d\theta_{e,0}}{dz}    \right\rangle 
    = \,\, \underbrace{Q_{s,n}}_{\text{shallow}} 
    \quad + 
    \,\,
    \underbrace{Q_{d,n}}_{\text{deep}}.
\end{equation}
For simplicity, we parameterize $Q_{s,n}$ and $Q_{d,n}$ as piecewise-constant functions:
\begin{equation}
\begin{split}
    Q_{s,n}=\begin{cases}
        <0\,\,\text{const at the shallow stage},\\
        =0,\,\,\text{at the deep stage},
    \end{cases}
    \quad
    Q_{d,n}=\begin{cases}
        =0\,\,\text{at the shallow stage},\\
        <0,\,\,\text{const at the deep stage},
    \end{cases}
\end{split}   
\end{equation}
with $|Q_{s,n}|<|Q_{d,n}|$. Here comes our key assumption: The system alternates between shallow and deep convection stages. During the shallow convection stage, shallow cumulus and congestus clouds form, while in the deep convection stage, deep convective clouds develop, which can detrain and lead to the formation of stratiform clouds. See Fig. \ref{fig:cartoon_dual_threshold} for an illustration. The convection term is negative (see Appendix B) because convection brings the free tropospheric dry air down to the boundary layer. Shallow convection and congestus convection bring down less dry air than deep convection and stratiform precipitation \cite[e.g.,][]{de2018variations}. At the shallow stage, surface heating dominates, raising $\Theta_n$. When $\Theta_n$ climbs up and reaches an upper threshold, by which the convective inhibition energy (CIN) is eliminated, the shallow stage transitions to the deep stage. When deep convection occurs, it consumes the convective available potential energy (CAPE) efficiently, decreasing $\Theta_n$. When $\Theta_n$ drops to the lower threshold, by which CAPE is used up, deep convection can no longer be sustained, and the deep stage transitions to the shallow stage. Figure \ref{fig:lag} shows that $\Theta_n$ is roughly in the opposite phase to CIN and the same phase to CAPE, suggesting that CIN is the smallest when $\Theta_n$ is the highest and CAPE is the lowest when $\Theta_n$ is the lowest. This piece of evidence supports their relevance to the upper and lower thresholds.\footnote{Strictly speaking, CIN and CAPE are functions of both $\theta_e$ in the boundary layer and buoyancy in the free troposphere \citep{emanuel1994book}. For future work, the influence of the free tropospheric buoyancy, which is influenced by the wave, could be represented by making the upper or lower threshold of $\Theta_n$ evolve with time. For example, if the lower troposphere has a cold anomaly, CIN will be reduced, lowering the upper threshold of $\Theta_n$. } The wave forcing influences the transition time but does not influence the amplitude of $\Theta_n$, which is prescribed by the lower and upper thresholds.



In the absence of wave forcing, the period of a convective lifecycle $T$ obeys:
\begin{equation}  \label{eq:T_Delta_Theta}
    T = \underbrace{ \frac{\Delta \Theta}{Q_{s,n} + Q_{rad} + Q_{surf} } }_{T_s} 
   \,\, + \,\, \underbrace{ \frac{\Delta \Theta}{-(Q_{d,n} + Q_{rad} + Q_{surf})} }_{T_d},
\end{equation}
Here $\Delta \Theta$ is the range of the lower and upper threshold of $\Theta$, essentially the amplitude of individual convective cell $A_{idv}$. The $T_s$ and $T_d$ denote the shallow and deep stage duration, respectively. We focus on the simplest case - symmetric shallow and deep stage duration ($T_s = T_d$). The $T_s \neq T_d$ case will be studied in Appendix C as an extension to the symmetric problem.

 
To quantitatively analyze the synchronization process, we can define the convective phase. The convective phase of the $n^{th}$ cell is defined as $\varphi_n$, which is a piecewise-linear function of $\Theta_n$:
\begin{equation} \label{eq:phase_definition}
    \varphi_n \equiv 
    \begin{cases}
        \pi \frac{\Theta_n}{\Delta \Theta} + \frac{\pi}{2}, \quad \text{shallow stage}, \\
        \frac{3}{2}\pi - \pi \frac{\Theta_n}{\Delta \Theta}, \quad \text{deep stage}. \\
    \end{cases}
\end{equation}
The $\varphi_n=0$ phase denotes the deep-to-shallow transition point, and the $\varphi_n=\pi$ phase denotes the shallow-to-deep transition point. Using Eqs. (\ref{eq:T_Delta_Theta}) and (\ref{eq:phase_definition}), we see that without the wave forcing, the convective phase increases at a constant angular velocity of $\omega$:
\begin{equation}  \label{eq:omega_expression}
    \omega \equiv 
    \begin{cases}
    \frac{\pi}{\Delta \Theta} \left( Q_{s,n} + Q_{rad} + Q_{surf} \right)
    = \frac{2\pi}{T},
    \quad \text{shallow stage}, \\
     -\frac{\pi}{\Delta \Theta} \left( Q_{d,n} + Q_{rad} + Q_{surf} \right)
    = \frac{2\pi}{T},
    \quad \text{deep stage}.    
    \end{cases}
\end{equation}
We name $\omega$ the intrinsic convective angular frequency.

\subsection{Simulation result of the microscopic model}\label{subsec:microscopic_result}

We perform a simulation of the microscopic model with an ensemble member size of $N=100$,
forced by a cosine-shape function that mimics the gravity wave:
\begin{equation}  \label{eq:wave_forcing}
    - \widetilde{W} \left\langle \frac{d\theta_{e,0}}{dz} \right\rangle
    = - \widetilde{W}_0 \left\langle \frac{d\theta_{e,0}}{dz} \right\rangle \cos(\Omega t).
\end{equation}
Here, $\widetilde{W}_0$ is a fixed amplitude of the wave's vertical velocity, and $\Omega$ is the wave's angular frequency. We study the simplest case:
\begin{itemize}
    \item The shallow stage and deep stage duration times are equal ($T_s=T_d$).
    \item The convective intrinsic angular frequency and the wave angular frequency are equal ($\omega = \Omega$).
\end{itemize}
The simulation is initiated by uniformly prescribing the convective phase over the members. Each convective cell is independent, except that they are coordinated by the wave forcing. Substituting Eqs. (\ref{eq:Fsn_Fdn}) and (\ref{eq:wave_forcing}) into Eq. (\ref{eq:Theta_simplified}), the governing equation of each cell's $\Theta_n$ is shown below:
\begin{equation}
    \frac{d\Theta_n}{dt} = Q_{s,n} + Q_{d,n} + Q_{rad} + Q_{surf} - \widetilde{W}_0 \left\langle \frac{d\theta_{e,0}}{dz} \right\rangle \cos(\Omega t).
\end{equation}

There are four input parameters: $\Delta \Theta$, $Q_{s,n}+Q_{rad}+Q_{surf}$ [equal to $-(Q_{d,n}+Q_{rad}+Q_{surf})$], $-\widetilde{W}_0 \langle d\theta_{e,0}/dz \rangle$, and $\Omega$. The matching of wave and convective frequency,
\begin{equation}
    \frac{2\pi}{\Omega} = \frac{2 \Delta \Theta}{Q_{s,n}+Q_{rad}+Q_{surf}},
\end{equation}
provides a constraint. Viewing $\Theta_n / \Delta \Theta$ as the nondimensional equivalent potential temperature perturbation and $\Omega t$ as the nondimensional time, it appears that the system has only one nondimensional parameter, which is named $B$:
\begin{equation}  \label{eq:B}
    B \equiv 
    - \frac{ \widetilde{W}_0 \left\langle \frac{d\theta_{e,0}}{dz} \right\rangle }{Q_{s,n}+Q_{rad}+Q_{surf}}
    = - \frac{\pi \widetilde{W}_0 }{\Omega \Delta \Theta} \left\langle \frac{d\theta_{e,0}}{dz} \right\rangle.
\end{equation}
The parameter $B$ measures the strength of wave forcing on convection. Using $\widetilde{W}\sim 10^{-3}$ m s$^{-1}$ (at $t=5$ days, Fig. \ref{fig:synchro_time_series}d), $\langle d\theta_{e,0}/dz \rangle \approx -6$ K km$^{-1}$ in the boundary layer (Fig. \ref{fig:the_profile}a), $\Omega \approx 2 \pi$ day$^{-1}$ (Fig. \ref{fig:synchro_time_series}), we get $B\approx 0.26$. We have tried one simulation with this setting, but the synchronization is too fast compared to the cloud-permitting simulation. We leave the puzzle aside temporarily and adopt a weaker forcing of $B=0.075$ to explore the property of the microscopic model. The wave and convective period are set to 1 day, and $\Delta \Theta$ is set to 1 K. 





\begin{figure}[h]
\centerline{\includegraphics[width=1.4\linewidth]{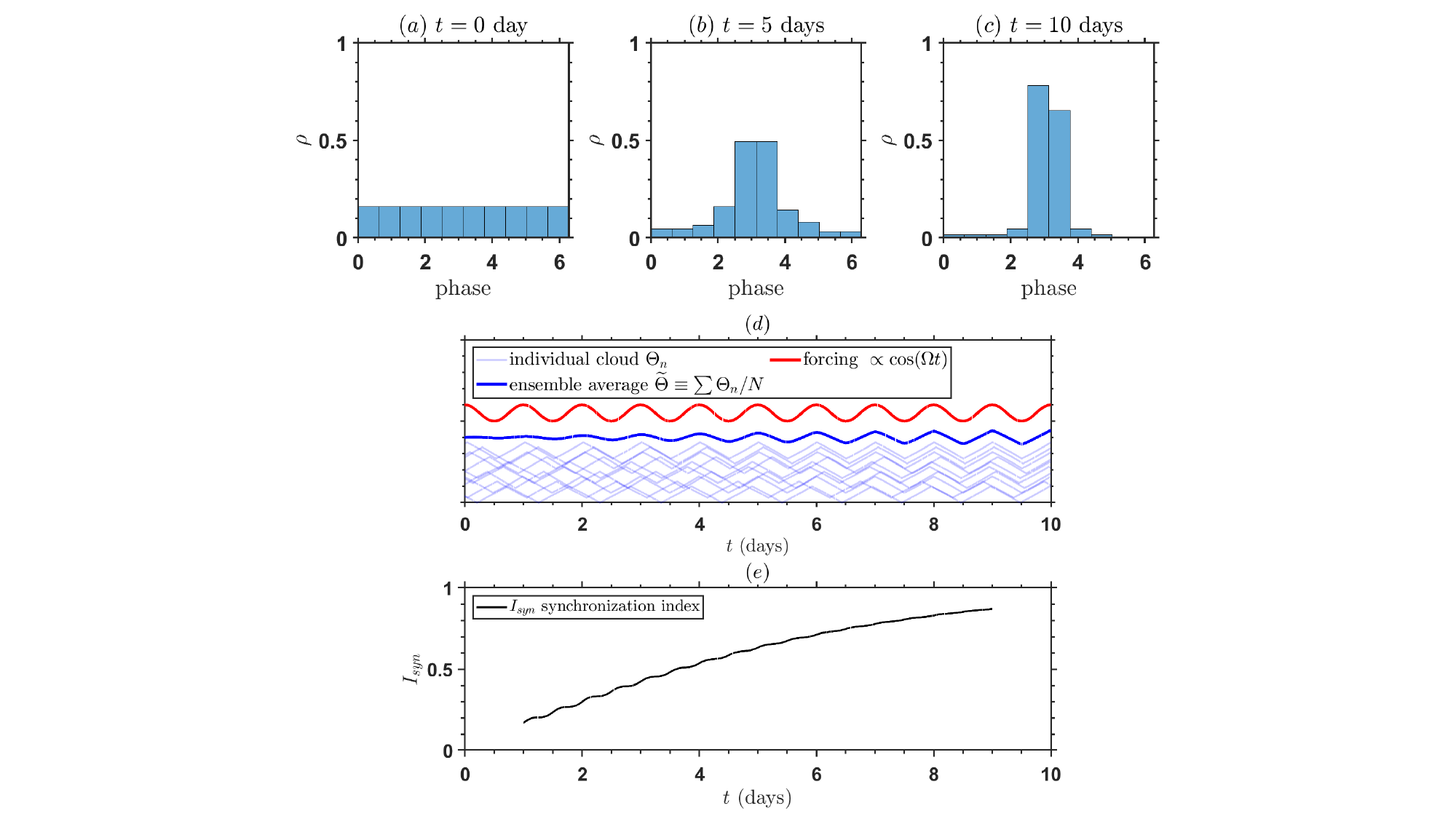}}
\caption{The simulation result of the microscopic model with 100 members, using $B=0.075$. The numerical scheme uses Euler forward with a time step of 60 seconds. (a), (b) and (c) are histograms of the convective lifecycle phase at $t=0$, $t=5$ days, and $t=10$ days. The horizontal axis is from 0 to $2\pi$. The deep-to-shallow transition phase is $\varphi=0$, and the shallow-to-deep transition phase is $\varphi=\pi$. (d) shows the time series. The y-axis scale of the time series is removed to facilitate the comparison of phases. The light blue lines show the $\Theta_n$ of 10 randomly sampled members. The dark blue line shows the ensemble average of all members, $\widetilde{\Theta}$. The red line shows the wave forcing $-\widetilde{W} \langle d \theta_{e,0}/dz \rangle$. (e) shows the synchronization index $I_{syn}$, calculated with a two-day-long moving time window.  }   \label{fig:synchro_show_hist}
\end{figure}   

Figure \ref{fig:synchro_show_hist} shows that the convective members gradually synchronize to the wave forcing, analogous to the cloud-permitting simulation result shown in Fig. \ref{fig:synchro_time_series}. Meanwhile, the amplitude of the ensemble average $\widetilde{\Theta}$ grows, accompanied by the rise of the synchronization index $I_{syn}$. As the amplitude of $\Theta$ grows, its shape gradually transitions from a sinusoidal to a tent shape (Fig. \ref{fig:synchro_show_hist}d), marking the completion of synchronization. 

\begin{figure}[h]
\centerline{\includegraphics[width=1.2\linewidth]{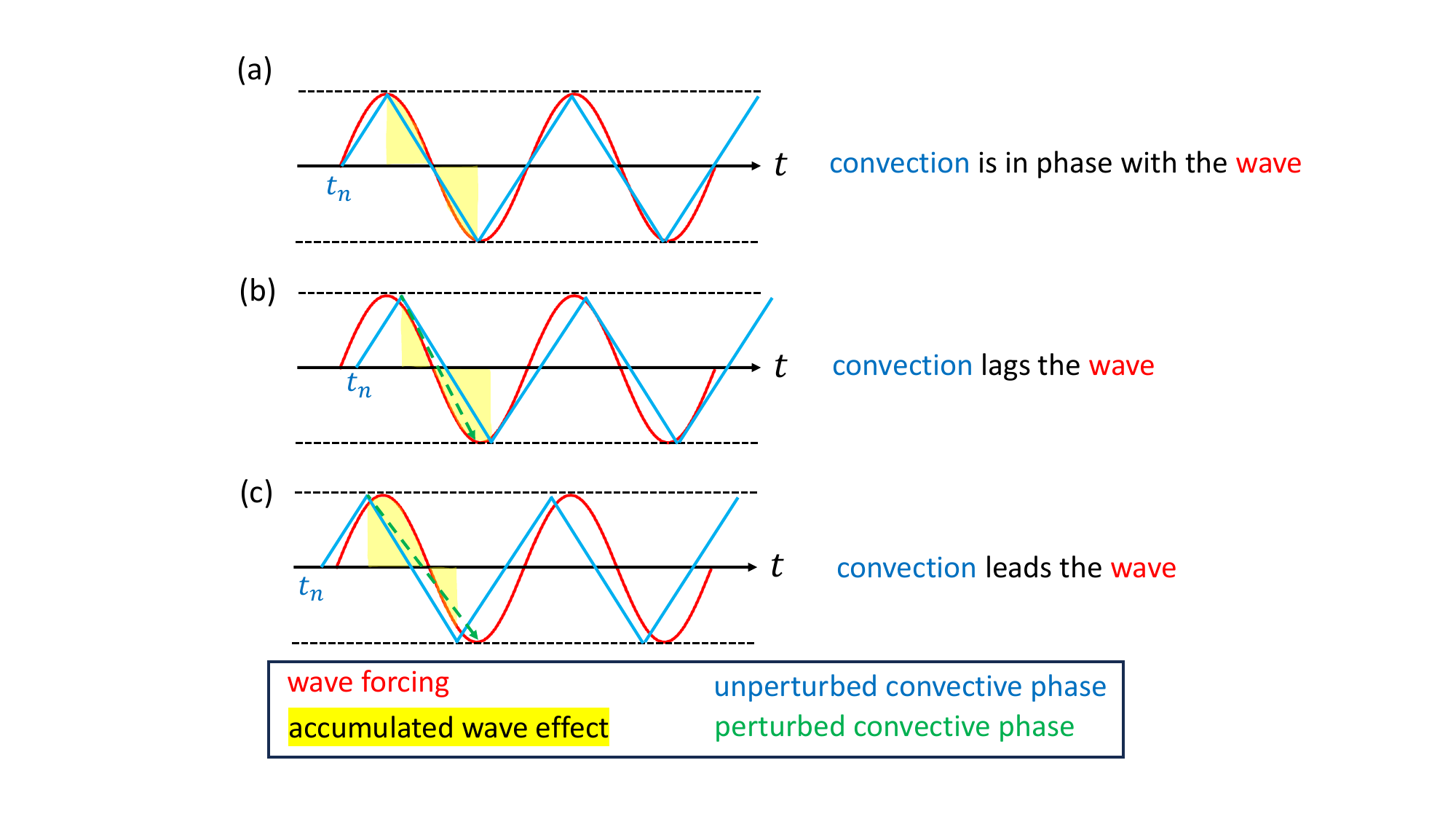}}
\caption{A schematic diagram of the adjustment of the convective phase by an external wave forcing in the symmetric case ($T_s=T_d$). The red line sketches the wave forcing $-\widetilde{W} \langle d\theta_{e,0}/dz \rangle$. The blue line sketches the evolution of $\Theta_n$ without the wave perturbation. The dashed green arrow sketches the $\Theta_n$ perturbed by the wave. The yellow shading sketches the accumulated effect of the wave's influence on the convective phase within the deep stage, which is proportional to $\int_{t_n + T/2}^{t_n + T} -\widetilde{W} \langle d\theta_{e,0}/dz \rangle dt$. The blue symbol $t_n$ marks the deep-to-shallow transition time. (a) shows the case where convection is in phase with the wave; (b) shows the case where convection lags the wave; (c) shows the case where convection leads the wave. The governing equation of the phase adjustment process is shown in Eq. (\ref{eq:phase_drift_main}).}   \label{fig:cartoon_adjustment}
\end{figure}

We use Fig. \ref{fig:cartoon_adjustment} to explain the mechanism of the phase drift by taking one convective lifecycle as an example. The lifecycle starts from $t_n$ and ends at $t_n + T$. The shallow stage is $t_n<t<t_n + T/2$, and the deep stage is $t_n + T/2 < t < t_n + T$. Both stages contribute to the phase drift. We use the deep stage to explain the physical process. 
\begin{itemize}
    \item When convection is in phase with the wave (Fig. \ref{fig:cartoon_adjustment}a), it is a locked state. This is because the accumulated wave forcing at the deep stage, represented by $\int_{t_n + T/2}^{t_n + T} -\widetilde{W} \langle d\theta_{e,0}/dz\rangle dt$, is zero. 
    \item When convection lags the wave (Fig. \ref{fig:cartoon_adjustment}b), the wave forcing accelerates the lowering of $\Theta_n$ at the deep stage, making it end earlier and adjusting the convective phase towards the wave. 
    \item When convection leads the wave (Fig. \ref{fig:cartoon_adjustment}c), the wave forcing decelerates the lowering of $\Theta_n$ at the deep stage, making it end later and also adjusting the convective phase towards the wave.
\end{itemize}
A similar analysis can be performed for the shallow stage, which yields the same result. In summary, the wave synchronizes the convective lifecycle by modulating its phase and adjusting it to a locked state.
At the locked state, the wave forcing and the ensemble-averaged quantity $\widetilde{\Theta}$ are in phase, agreeing with the cloud-permitting simulation (Fig. \ref{fig:synchro_time_series}d).

Note that we need to use an unrealistically weak forcing of $B=0.075$ to make the microscopic model qualitatively match the cloud-permitting simulation. As Appendix C confirms, at least two missing factors may suppress the synchronization: the asymmetry between shallow and deep stages' duration time ($T_s \neq T_d$) and noise on $\Theta_n$. We keep these limitations in mind and move forward to explore the collective effect of synchronization by deriving a governing equation of the macroscopic variable $\widetilde{\Theta}$.

\subsection{The harmonic oscillator approximation}\label{subsec:oscillator}

In this subsection, we theorize the qualitative understanding from the simulation and report the governing equation of $\widetilde{\Theta}$. The detailed derivation procedure is documented in Appendix D, and the main idea is sketched here. First, we follow the intuition revealed in Fig. \ref{fig:cartoon_adjustment} to derive the phase drift equation of an individual convective cell. When the wave and convection frequency equals ($\Omega=\omega$), the phase drift equation [Eq. (\ref{eq:theta_n_equation})] reads: 
\begin{equation}  \label{eq:phase_drift_main}
    \frac{d\varphi_n}{dt}
    \approx 
    \omega + \frac{2 \widetilde{W}_0}{\Delta \Theta} \left\langle \frac{d\theta_{e,0}}{dz} \right\rangle \sin \left( \varphi_n - \varphi_w \right),
\end{equation}
where $\varphi_w \equiv -\pi + \Omega t$ is the wave phase. It shows that $\varphi_n$ adjusts towards $\varphi_w$ with a timescale of $\tau$:
\begin{equation}
    \tau \equiv - \frac{\Delta \Theta}{2 \widetilde{W_0}} \left\langle \frac{d\theta_{e,0}}{dz} \right\rangle^{-1} = \frac{\pi}{2B} \Omega^{-1}.
\end{equation}
Here, we have used the definition of $B$ from Eq. (\ref{eq:B}). The $\tau$ is fundamentally related to the convective period and the forcing magnitude, with a shorter period or stronger forcing accelerating the synchronization. Second, we derive the governing equation of the phase distribution function $\rho$. It obeys a conservation law, with the phase drifting rate as the transport flux between phase bins. Third, we obtain an analytical solution of $\rho$ and further solve $\widetilde{\Theta}$, which approximately obeys a forced harmonic oscillator equation:
\begin{equation} \label{eq:oscillator}
    \frac{d^2 \widetilde{\Theta}}{dt^2} + \omega^2 \widetilde{\Theta} 
    \approx - \frac{8}{\pi^2} \frac{d}{dt}\left( \widetilde{W} \left\langle \frac{d\theta_{e,0}}{dz} \right\rangle \right).
\end{equation}
Here, $\omega$ is the intrinsic angular frequency of convection. The oscillator approximation is valid under four assumptions:
\begin{enumerate}
    \item The shallow and deep stage duration times equal each other [$T_s=T_d$].
    \item The wave forcing period is not far from the convective lifecycle period [$|\omega - \Omega| \ll \omega$]. 
    \item The synchronization status is weak [the ensemble-averaged quantity being much smaller than the fluctuation magnitude of an individual cloud $\widetilde{\Theta} \ll \Delta \Theta$].
    \item The wave forcing is weak [$B \ll 1$].        
\end{enumerate}
Among them, Assumption 1 is for simplicity, with the $T_s \neq T_d$ case not yet analytically tractable; Assumptions 2 and 3 are used in deriving the phase drift equation (\ref{eq:phase_drift_main}); Assumptions 3 and 4 are used in linearizing the phase distribution ($\rho$) equation, a key step in deriving the oscillator equation (\ref{eq:oscillator}) from the phase drift equation (\ref{eq:phase_drift_main}).

\begin{figure}[h]
\centerline{\includegraphics[width=1.1\linewidth]{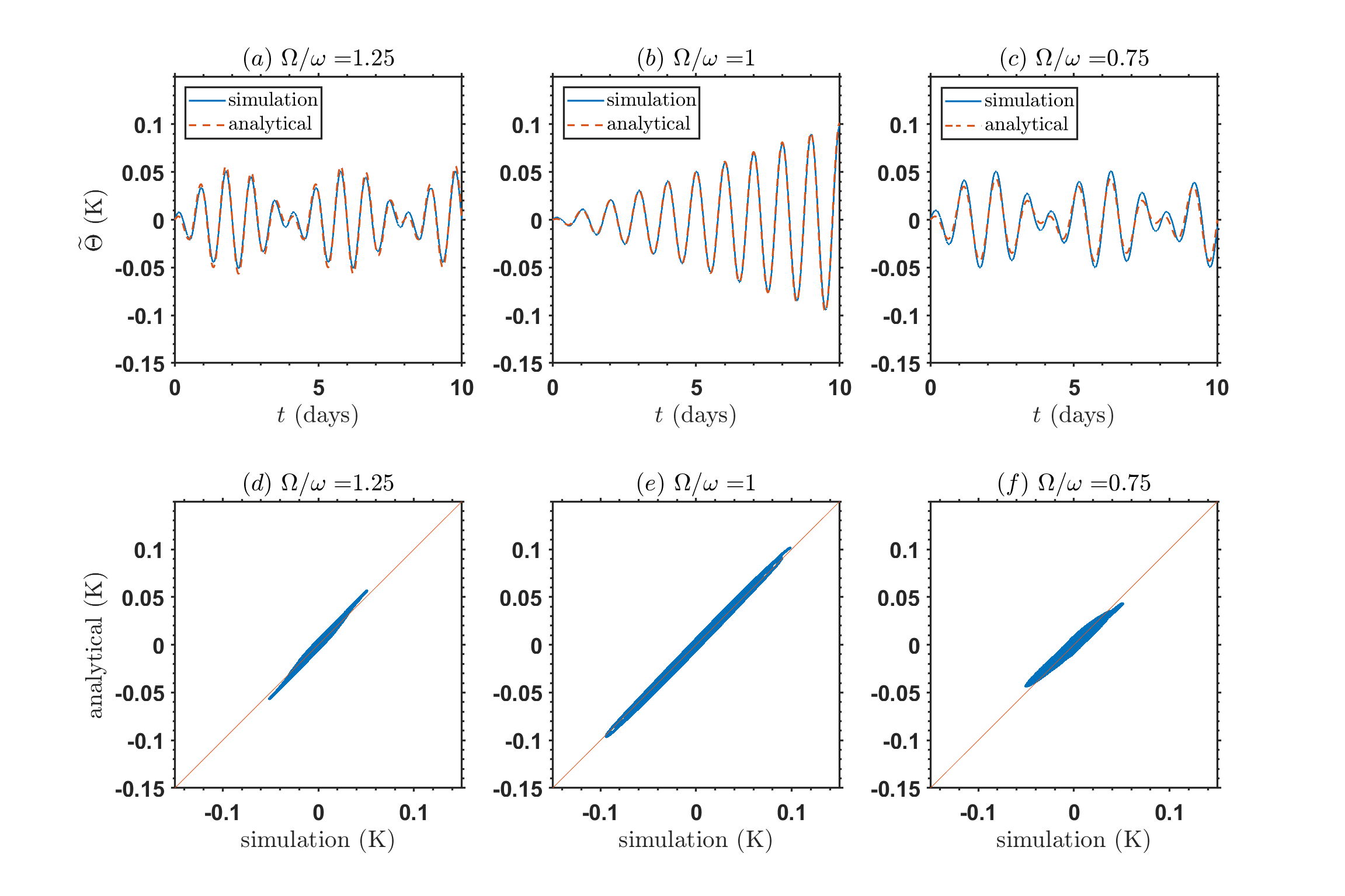}}
\caption{A validation of the oscillation approximation of the cloud ensemble [Eq. (\ref{eq:oscillator})]. (a)-(c) shows the numerical simulation of the microscopic model (solid blue line) and the analytical solution [Eqs. (\ref{eq:analytical_A}) and (\ref{eq:oscillator_solution_neq}), dashed red line] of three cases: $\Omega /\omega=1.25$, 1, and 0.75. Here, $\Omega/\omega$ is the ratio of wave angular frequency to the convective intrinsic angular frequency. The $\Delta \Theta$ is set to 1 K. The convective lifecycle period is fixed to be 1 day, and the time-dependent factor of the wave forcing is $\cos(\Omega t)$. The forcing amplitude $-\widetilde{W}_0 \langle d\theta_{e,0}/dz \rangle$ is $0.1$ K day$^{-1}$, $0.025$ K day$^{-1}$, and $0.1$ K day$^{-1}$ for the three cases. The motivation for using a weaker forcing for the resonant case ($\Omega/\omega=1$) is because its $\widetilde{\Theta}$ grows much faster than others. The second row compares the numerical and analytical solutions more closely. The solutions are plotted as blue dots; the red line is a 1:1 reference line.}   \label{fig:analytical_solution_test}
\end{figure}

Figure \ref{fig:analytical_solution_test} shows that the analytical solution [Eqs. (\ref{eq:analytical_A}) and (\ref{eq:oscillator_solution_neq})] under the oscillator approximation generally agrees with the numerical solution of the microscopic model for $\Omega/\omega = 1.25$, 1, and 0.75. 
When $\omega \neq \Omega$, but is not too far from each other, the cloud ensemble responds to the forcing as an envelope, reminiscent of the cloud-permitting simulations of \citet{davies2013departures}. When $\omega = \Omega$, the cloud ensemble resonates with the wave forcing, with the amplitude of $\widetilde{\Theta}$ growing linearly with time. A fundamental property of the resonant solution is $\widetilde{\Theta} \propto t \cos(\Omega t)$ [Eq. (\ref{eq:analytical_A})], which means $\widetilde{\Theta}$ and $\widetilde{W}$ are in phase. This in-phase relation agrees with the microscopic model (Fig. \ref{fig:synchro_show_hist}d) and the cloud-permitting simulation (Fig. \ref{fig:synchro_time_series}d). So far, this agreement is the most direct support to our synchronization model.

To understand the physics of the oscillator approximation, we heuristically derive Eq. (\ref{eq:oscillator}), starting from performing a derivative of the microscopic model equation $d\widetilde{\Theta}/dt$ [Eq. (\ref{eq:Theta_simplified})]:
\begin{equation}  \label{eq:oscillator_heuristic}
\begin{split}
    \frac{d^2 \widetilde{\Theta}}{dt^2} 
    &\approx \frac{1}{N}\sum_n \frac{d}{dt}\left( Q_{s,n} + Q_{d,n} + Q_{rad} + Q_{surf}  - \widetilde{W} \left\langle \frac{d\theta_{e,0}}{dz} \right\rangle \right) \\
    &\approx \frac{1}{N}\sum_n \frac{d}{dt}\left( Q_{s,n} + Q_{d,n} + Q_{rad} + Q_{surf} \right)
    - \frac{d}{dt}\left( \widetilde{W} \left\langle \frac{d\theta_{e,0}}{dz} \right\rangle \right).    
\end{split}    
\end{equation}
We use ``$\approx$" because the sum of the convection, radiation, and surface heating terms, $\left( Q_{s,n} + Q_{d,n} + Q_{rad} + Q_{surf} \right)$, is a piecewise-constant function of time, whose derivative is not well-defined in a regular sense. Its derivative is a pulse at the shallow-to-deep or deep-to-shallow transition points and takes zero elsewhere. Thus, the derivative depends on how many members are at the transition points, i.e., the phase distribution function $\rho$ at $\varphi=0$ and $\varphi=\pi$:
\begin{equation}\label{eq:restore}
\begin{split}
    \frac{1}{N} \sum_n \frac{d}{dt}\left( Q_{s,n} + Q_{d,n} + Q_{rad} + Q_{surf} \right) 
    \propto \underbrace{ \rho(\varphi=0) }_{\text{deep}\to\text{shallow}} - \underbrace{ \rho(\varphi=\pi)}_{\text{shallow}\to\text{deep}}
    \propto - \widetilde{\Theta}.
\end{split}
\end{equation}
The last step uses Eq. (\ref{eq:Theta_tilde_expression}), which links the shape of $\rho$ to the value of $\widetilde{\Theta}$. An intuitive way to understand this relation is by noticing that $\Theta_n$ reaches a valley at $\varphi=0$ (e.g., Fig. \ref{fig:cartoon_dual_threshold}), which contributes negatively to $\widetilde{\Theta}$, and reaches a peak at $\varphi=\pi$, which contributes positively to $\widetilde{\Theta}$. The above analysis shows that the restoring force of the oscillation is provided by the mutual transition between the shallow and deep stages of the convective lifecycle. 


\section{Discussion}\label{sec:discussion}

This section comments on the relationship between this microscopic model and previous studies, reviews the model's limitations, and sketches possible future research. 

The constructed dual-threshold system can be viewed as a nonlinear oscillator with a ``recharge-discharge" behavior. The slow growth and drop versus the rapid transition is analogous to the van der Pol oscillator \citep{van1926relax_oscill,strogatz2018nonlinear}, which describes a spring oscillator whose restoring force involves the cube of the displacement. However, we have found that the van der Pol oscillator does not manifest the synchronization behavior of this dual-threshold oscillator. Previous simple models have used nonlinear oscillators to depict the cloud ensemble, rather than the lifecycle of an individual cloud \cite[e.g.,][]{koren2011prey,yano2012ODEs,colin2021atmospheric}. 


The macroscopic oscillator model is derived under four assumptions: $T_s=T_d$, $|\omega-\Omega| \ll \omega$, $\widetilde{\Theta} \ll \Delta \Theta$, and $B\ll 1$. For the first assumption, we have not rigorously derived the $T_s \neq T_d$ equation, which has a thornier math. The numerical simulation in Appendix C shows that the synchronization still exists but is slower. We do not know whether the oscillator approximation is still valid. The third assumption is essentially a small-amplitude assumption. We need to extend it to the finite-amplitude state to depict the ``saturation" of synchronization. The fourth assumption is controversial because the magnitude of $B$ is estimated to be $B \sim 0.26$. We use $B \sim 0.075$ in the demonstrative calculations of the microscopic model, arguing that choosing a small $B$ effectively represents other neglected effects like noise and $T_s \neq T_d$. Changing $B$ is only a band-aid solution. We speculate that these omitted factors can be considered as damping effects. If so, our model would echo the prognostic cloud ensemble model of \citet{pan1998cumulus}, a linear damped oscillator. Their derivation is from the cloud ensemble's energetics, and our derivation is from the synchronization of convective lifecycles. The two derivations might be two different interpretations of the same subject.



A significant limitation of the current model is that the convective frequency is largely fixed, though the large-scale vertical advection of equivalent potential temperature can slightly influence it. A preliminary finding (not shown in the manuscript) is that the influence of lower tropospheric buoyancy on CIN can be parameterized as the fluctuation of the upper threshold of $\Theta_n$, which adjusts both the phase and frequency of convection. Another limitation is that clouds are not allowed to interact. For future work, one could consider cloud-cloud interaction due to cold pools \cite[e.g.,][]{feingold2013coupled} or the lateral exchange of water vapor \cite[e.g.,][]{craig2013coarsening,tan2018extended}, and investigate whether it positively or negatively contributes to cloud synchronization. It will be worth drawing analogies to the Kuramoto model \cite[e.g.,][]{strogatz2000kuramoto}, a paradigm model for the synchronization of a coupled oscillator that has been shown relevant to a variety of phenomena, including lightning \citep{yair2009clustering} and biochemical clocks \citep{zhang2020energy}. Last, one could couple the microscopic model with a shallow water equation to simulate how convection interacts with/via gravity waves and attempt to predict the growth rate of CCGW. 



Our highly idealized microscopic model demonstrates that a cloud ensemble might be a resonator via synchronization, made available by the two thresholds in shallow-to-deep and deep-to-shallow transitions. Previous literature has speculated that convection may resonate with gravity waves. \citet{bretherton2006interpretation} reported significant standing gravity waves when simulating a mock-Walker circulation in a rectangular domain using a nonuniform SST. They speculated that the wave resonates with convection. 
\citet{lane2011coupling} analyzed a 2D slab-symmetric cloud-permitting simulation, speculating that convection may resonate with a 13 m s$^{-1}$ gravity wave at the third baroclinic mode, which produces strong waves in the domain. \citet{yang2019convective} reported intense standing CCGWs in a 3D simulation with a doubly periodic square domain, uniform sea surface heat flux, and no sub-cloud rain evaporation. He inferred that the CCGWs are due to wave-convection resonance. Apart from gravity waves, \citet{bell2019mesoscale} found that the diurnal cycle may also interact with the convective lifecycle and modulate tropical cyclogenesis. They did not explicitly discuss whether the interaction is via resonance. Though enlightening, we have not seen any previous research that concretely answered whether clouds can resonate with waves. The complexities in both wave and convection make the question hard. As for the wave, a linear standing wave may not substantially differ from a traveling wave, but not for the nonlinear regime. Our budget analysis in Appendix B shows the background tendency term $-d\langle \theta_{e,0} \rangle/dt$ is significant at the late stage and has twice the frequency of the perturbation quantity $\widetilde{\Theta}$, indicating it is a super-harmonic mode nonlinearly generated by the standing wave. We must be cautious about this super-harmonic mode when interpreting any finite-domain simulations. As for convection, a cloud ensemble might consist of different cloud types with different characteristic frequencies. Thus, we boldly speculate that resonance might be too ubiquitous to identify, like the air we breathe. This paper shows a possible microscopic indicator of resonance: synchronization. Synchronization is equivalent to resonance in our microscopic model, but how related they are in the real atmosphere is unclear. Despite this, our amplitude decomposition algorithm can conveniently diagnose synchronization from data. It might provide an indirect path to test the wave-convection resonance hypothesis. We iterate that ``resonance" is a terminology that must be used cautiously.


\section{Conclusion}\label{sec:conclusion}

This paper investigates the synchronization of convective lifecycles, a collective behavior of tropical convection that has not received much attention. First, we use a coarse-graining technique to diagnose convective synchronization in a cloud-permitting RCE simulation. The horizontal anomaly of the boundary layer equivalent potential temperature averaged in 20 km $\times$ 20 km coarse-grained cells ($\Theta_n$) is used to represent the lifecycle of the $n^{th}$ cell, with $n=1$, 2, 3, ... We devised an amplitude decomposition algorithm, which shows that the response of a cloud ensemble to gravity waves is contributed by both the amplitude growth of individual clouds and their synchronization. The latter can be quantified with a synchronization index $I_{syn}$.

To understand the synchronization process, we build a simple microscopic model of individual clouds governed by the variable $\Theta_n$. The evolution of $\Theta_n$ is driven by convection, radiative cooling, and surface heating and is modulated by gravity waves. The radiation is assumed to be a constant cooling effect, and the surface flux is a constant heating effect. The convection term always reduces $\Theta_n$ but has different strengths at different stages. At the shallow stage, where most clouds are shallow cumuli and congestus, the convection term is set to be an insignificantly negative value, and $\Theta_n$ gradually rises. When $\Theta_n$ reaches the upper threshold, i.e., CIN is eliminated, the system transitions to the deep stage, where most clouds are deep convection and stratiform clouds. The convection term is set to be a more negative value, driving $\Theta_n$ to drop. When $\Theta_n$ drops to the lower threshold, i.e., CAPE is exhausted, deep convection cannot self-sustain, and the system transitions to the shallow stage. The wave, parameterized as a periodic forcing, modulates $\Theta_n$ via the vertical advection of the background equivalent potential temperature in the boundary layer. The wave can influence the arrival time of $\Theta_n$ to the transition thresholds and, therefore, modulate its phase. 

The clouds in the microscopic model show an intriguing collective behavior. When the convective period equals the wave period, members in the cloud ensemble are gradually synchronized. The synchronization is explained as a phase adjustment process. For example, when convection lags the wave, the wave accelerates convection to make it catch the wave, adjusting it to a locked state where $\Theta_n$ is in phase with the wave vertical velocity $\widetilde{W}$. A quantitative analysis of the phase adjustment is performed by first deriving a model of individual clouds' phase $\varphi_n$, then making a model of the phase distribution function $\rho$, and finally using $\rho$ to obtain an analytical solution of the ensemble-averaged quantity $\widetilde{\Theta}$, which approximately obeys a harmonic oscillator. The harmonic oscillator approximation is valid under four assumptions: i) an equal time duration of the shallow and deep stage, ii) a small frequency difference between the wave and convection, iii) a weak synchronization status, and iv) a weak wave forcing. When the convection and wave frequency match, the ensemble quantity $\widetilde{\Theta}$ exhibits a resonant growth. The analytical solution reproduces the in-phase relation between $\widetilde{\Theta}$ and $\widetilde{W}$ seen in the cloud-permitting simulation. The key findings within the microscopic model are summarized below:
\begin{enumerate}
    \item The dynamics are nonlinear for individual clouds but are approximately linear for the cloud ensemble.  
    \item The phase modulation of individual clouds appears as the amplitude modulation of the cloud ensemble.
    \item The restoring force of the cloud ensemble oscillation is provided by the shallow-to-deep and deep-to-shallow stage transition.  
    \item The synchronization is essentially the resonance of the cloud ensemble with the external forcing.
\end{enumerate}


The microscopic model suggests that synchronization is a potential way for a cloud ensemble to respond to an oscillatory forcing, when the wave period is close to the convective lifecycle. We hope the theory of synchronization and the algorithm for diagnosing synchronization can advance the research of tropical convection and be applied to improve convective parameterization schemes.

\acknowledgments
 Hao Fu is supported by the T. C. Chamberlin Fellowship from the University of Chicago. Da Yang is supported by the Packard Fellowship for Science and Engineering and an NSF CAREER Award (AGS-2048268). The initial idea of this paper emerged at the 2022 Boulder Summer School for Condensed Matter Physics, where Yuhai Tu gave a lecture on the synchronization of biochemical clocks. We thank Bolei Yang for a helpful discussion that reminded us of the importance of phase in understanding convection-wave coupling. He also provided valuable feedback on the synchronization index and proofread the manuscript. We thank Chris Li, Zhihong Tan, George Kiladis, Morgan E. O'Neill, Zhaohua Wu, and Yong-Qiang Sun for enlightening conversations. We also thank George Bryan for developing and maintaining the CM1 model, Stanford Research Computing Center, the CISL Lab of NSF NCAR, and the Research Computing Center of the University of Chicago for providing computational resources.

%
%
\datastatement
The movie version of Figs. \ref{fig:2Dxy_OLR} and \ref{fig:checkerboard}, the microscopic model, the data analysis codes, and a math derivation note are deposited in the supplemental material at Zenodo with a doi: 10.5281/zenodo.14164798. The simulation data is available by contacting the first author.


\appendix[A]
\appendixtitle{Amplitude decomposition}\label{subsec:ampliude_decomposition}

This appendix aims to prove the amplitude decomposition relation [Eq. (\ref{eq:wave_decomposition})], which decomposes the cloud ensemble amplitude into the mean amplitude growth of individual cells and the contribution from synchronization. Consider a 2-D array $\Theta_{np}$ that depicts the cloud ensemble, with $n=1$, 2, ..., $N$ denoting different cloud member and $p=1$, 2, ..., $P$ denoting the time snapshots. To better capture the signal, the time averaging window is advised to be roughly an integral number of the wave period (here taken as two days, approximately two wave periods). We will frequently use $\widetilde{\Theta}$ as the ensemble average:
\begin{equation}
\label{eq:ensemble_average}
    \widetilde{\Theta}_p \equiv \frac{1}{N} \sum_n \Theta_{np}.
\end{equation}
Next, we substitute Eq. (\ref{eq:ensemble_average}) into the definition of the cloud ensemble amplitude $A_{ens}$, individual convective amplitude $A_{idv}$, and synchronization index $I_{syn}$ [Eqs. (\ref{eq:A_wave})-(\ref{eq:Isyn})] to find their relationship.

For $A_{ens}$, we have:
\begin{equation}  \label{eq:A_wave_simplify}
\begin{split}
    A^2_{ens} 
    &= \frac{1}{P} \sum_p \left( \frac{1}{N}\sum_n \Theta_{np} \right)^2 \\
    &= \frac{1}{P} \sum_p \widetilde{\Theta_p}^2,
\end{split}
\end{equation}

For the product of $A_{idv}$ and $I_{syn}$, we have:
\begin{equation}  \label{eq:idv_syn_product}
\begin{split}
    A^2_{idv} \times I^2_{syn}
    &\approx \frac{1}{NP} \sum_p \sum_n \Theta_{np}^2
    \left[ 1 - \frac{ \frac{1}{NP} \sum_p \sum_n \left( \Theta_{np} - \widetilde{\Theta}_p \right)^2 }{ \frac{1}{NP} \sum_p \sum_n \Theta_{np}^2 }  \right] \\    
    &=  \frac{1}{NP} \sum_p \sum_n \Theta_{np}^2 - \frac{1}{NP} \sum_p \sum_n \left( \Theta_{np} - \widetilde{\Theta}_p \right)^2 \\
    &= \frac{1}{NP} \sum_p \sum_n \Theta_{np}^2 - \frac{1}{NP} \sum_p \sum_n \Theta_{np}^2 - \frac{1}{NP} \sum_p \sum_n \widetilde{\Theta_{p}}^2
    + \frac{2}{NP} \sum_p \sum_n \Theta_{np} \widetilde{\Theta_p} \\
    &= \frac{1}{P} \sum_p \widetilde{\Theta_p}^2.
\end{split}    
\end{equation}
Eqs. (\ref{eq:A_wave_simplify}) and (\ref{eq:idv_syn_product}) have the same result, proving Eq. (\ref{eq:wave_decomposition}).

\appendix[B]
\appendixtitle{Budget of the boundary layer equivalent potential temperature}\label{subsec:budget}

In this appendix, we perform a budget analysis of $\theta_e$ in the boundary layer.

\begin{figure}[h] \centerline{\includegraphics[width=0.6\linewidth]{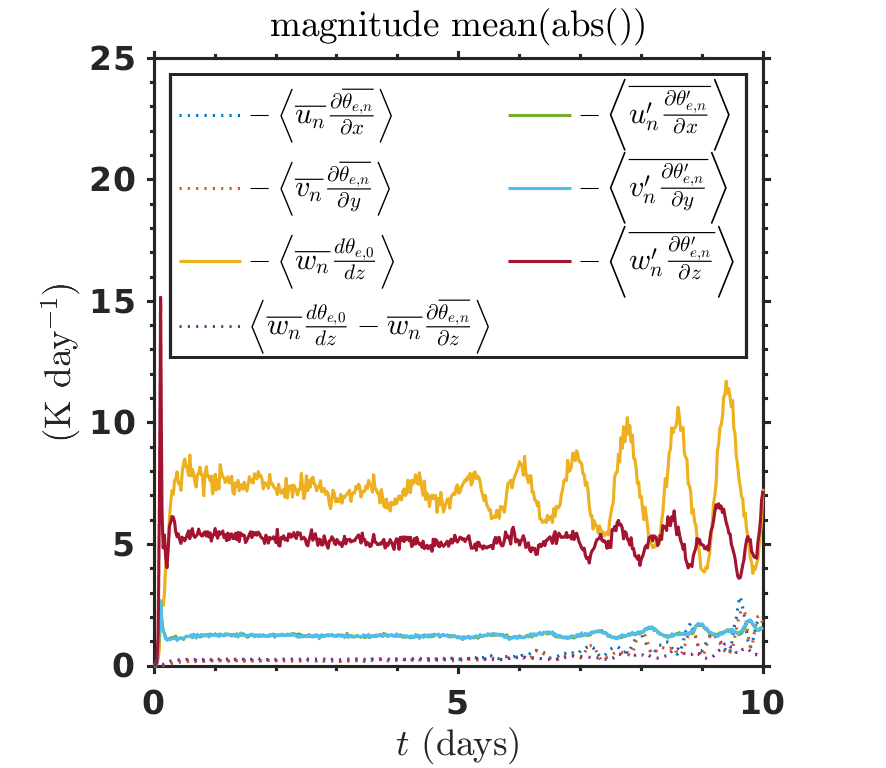}}
\caption{The magnitude of the advective term in the $\Theta$ equation decomposed into the mean parts and eddy parts [Eq. (\ref{eq:Theta_decompose})], diagnosed from the CM1 simulation. See the legend for the meaning of each term. The magnitude is represented by calculating each term's absolute value and then its ensemble average. The terms to be neglected in the microscopic model are marked with dotted lines. Other terms are shown as solid lines. The phase information in this figure is hard to interpret and not quite useful, due to the application of the absolute value operator. 
}   \label{fig:std_magnitude_compare}
\end{figure}

First, we justify some approximations in deriving the microscopic model, shown in Eqs. (\ref{eq:assumptions_Theta}) and (\ref{eq:assumptions_Theta_horizontal}). Figure \ref{fig:std_magnitude_compare} shows the magnitude of the decomposed advection terms in the original $\Theta$ equation [Eq. (\ref{eq:Theta_decompose})]. The $\partial/\partial x$, $\partial/\partial y$, and $\partial/\partial z$ are calculated with second-order central differences, except at the bottom where $\partial/\partial z$ is calculated with the first-order difference. The dotted lines show the terms to neglect, which are much smaller than others.

\begin{figure}[h] \centerline{\includegraphics[width=1\linewidth]{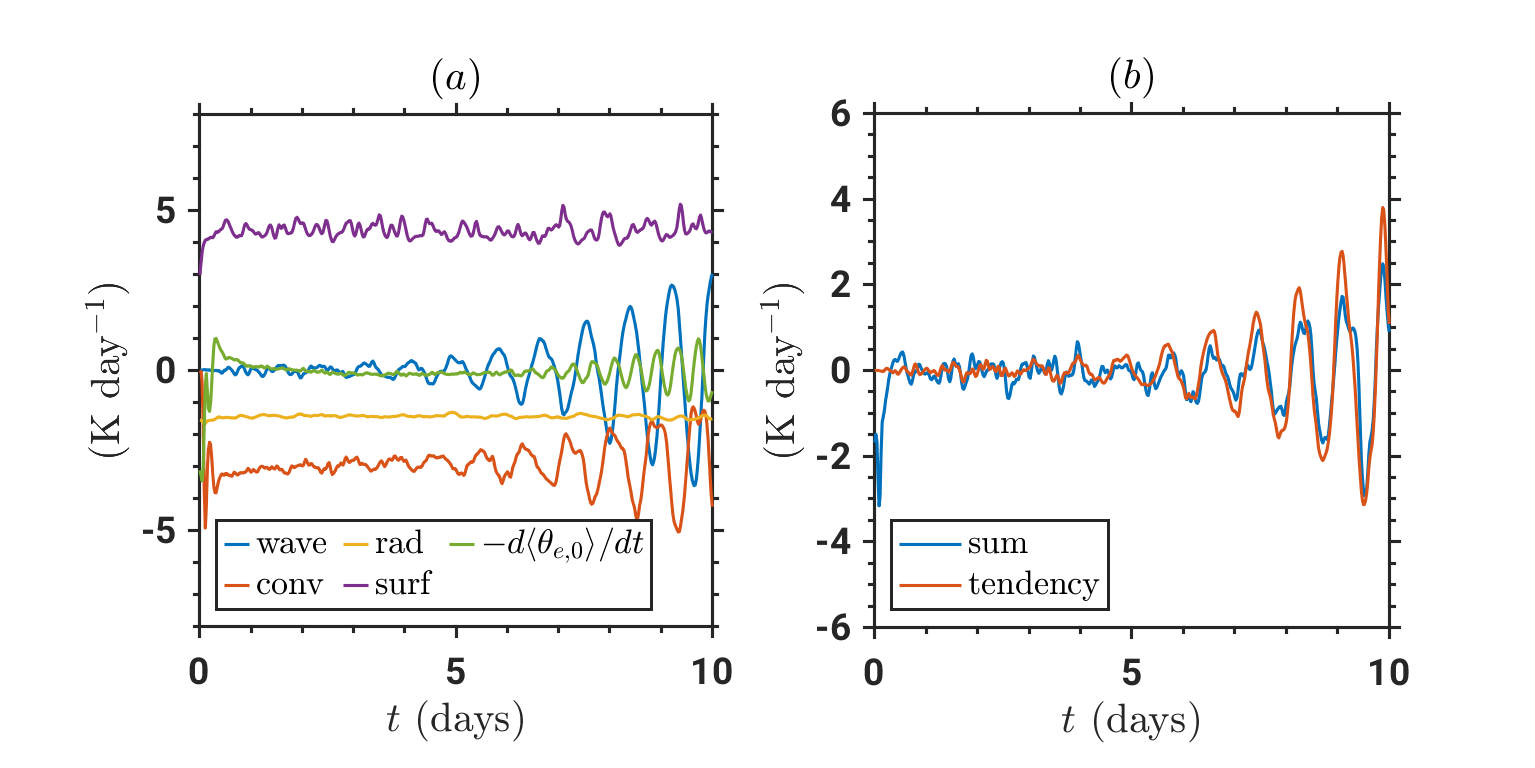}}
\caption{Budget analysis of the $\Theta_n$ equation (\ref{eq:Theta_simplified}) after an ensemble average. (a) The ensemble-averaged wave term (blue line), convection term (red line), radiation term (yellow line), surface heating term (purple line), and the background tendency term $-d\langle \theta_{e,0} \rangle/dt$ (green line). (b) The sum of the five terms (blue line) and the tendency term $d\widetilde{\Theta}/dt$ (red line). The tendency terms are calculated with the second-order central difference in time, except for the first and last snapshots where the first-order difference is used. To ease visualization, all the time series have been smoothed by a 3-hour-scale Gaussian filter.
}   \label{fig:std_wave_convection}
\end{figure}

Second, we compare the time evolution of the combined terms in Eq. (\ref{eq:Theta_simplified}): wave term, convection term, radiation term, surface heating term, and the background tendency term $-d\langle \theta_{e,0} \rangle/dt$. The result calculated for individual convective cells is too noisy to identify useful information, so we plot their ensemble average in the boxed region in Fig. \ref{fig:std_wave_convection}a. This indirect diagnosis only reveals an individual cell's property at the late stage by which the synchronization is significant. The wave term gradually amplifies, as expected. The convection term has an amplifying trend with a base value of around $-3$ K day$^{-1}$. The radiation term, which includes both shortwave and longwave radiation, is around a constant near $-1.5$ K day$^{-1}$. The surface heating term is around a constant near $4.5$ K day$^{-1}$. The background tendency term $-d\langle \theta_{e,0} \rangle/dt$ is minor. It has almost twice the frequency of convection and wave terms, indicating its nonlinear origin. The subgrid fluxes, which is only significant in the mixed layer, i.e., the lowest 0.5 km, is supposed to be an interior mixer within the boundary layer and is omitted. Figure \ref{fig:std_wave_convection}b shows that the budget is roughly closed. 




\appendix[C]
\appendixtitle{Asymmetry and noise}\label{subsec:asymmetry}

\section{Asymmetric shallow and deep stages}

The assumption that the shallow stage duration time equals the deep stage duration time ($T_s=T_d$) is mathematically convenient. However, in the real atmosphere, the deep convective time is usually much shorter than the shallow convective time, if not counting the stratiform precipitation time. Thus, we must investigate the more general case where $T_s \neq T_d$.

\begin{figure}[h]
\centerline{\includegraphics[width=1.2\linewidth]{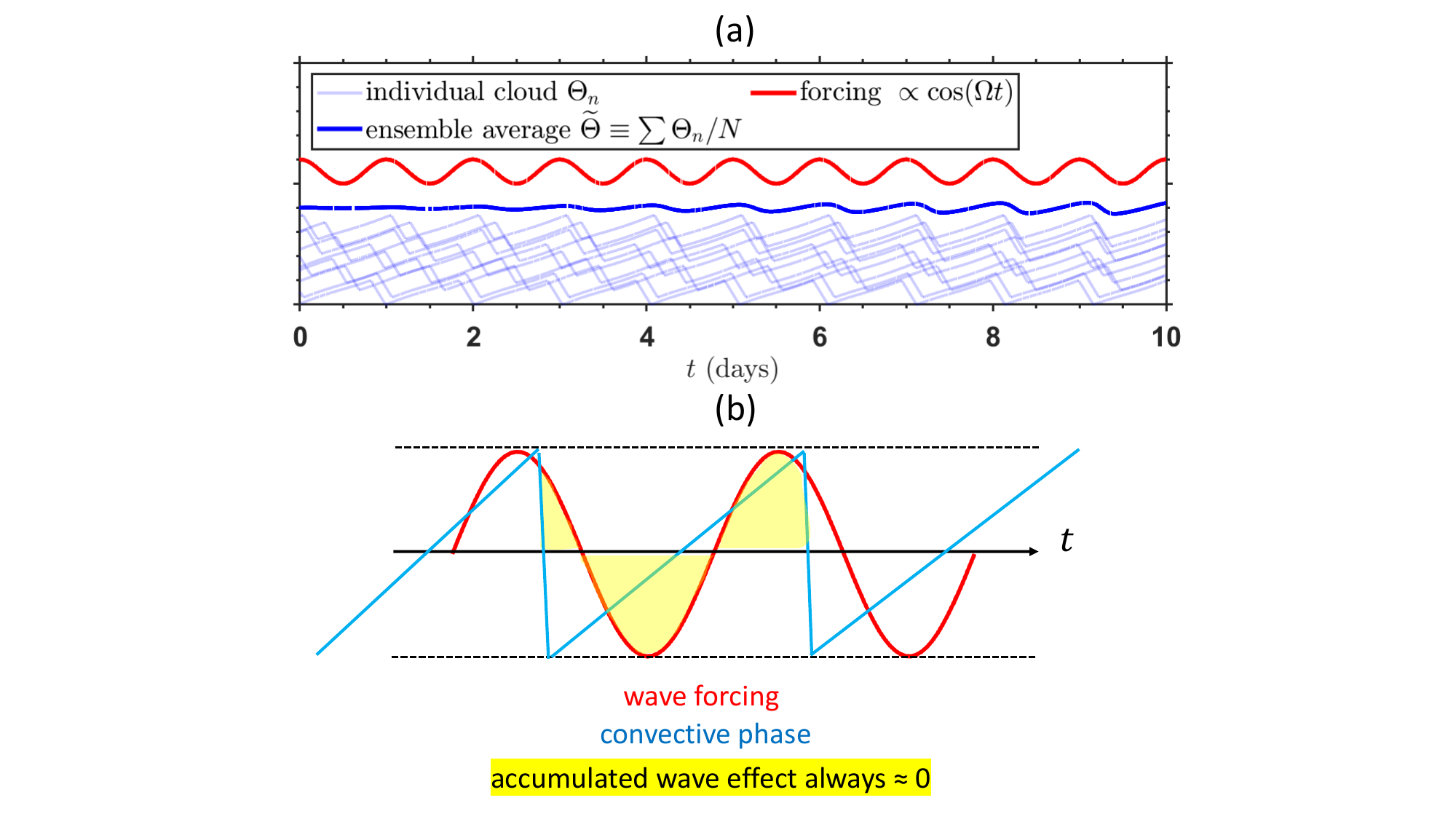}}
\caption{(a) The same as Fig. \ref{fig:synchro_show_hist}d, but for an illustration of the synchronization process of the asymmetric $T_s/T_d=5$ case. The synchronization is much slower than the $T_s/T_d=1$ case. (b) A schematic diagram of the adjustment of the convective phase by an external wave forcing in an extremely asymmetric case ($T_s \gg T_d$). The legend follows Fig. \ref{fig:cartoon_adjustment}. }   \label{fig:synchro_asymmetry_run=4}
\end{figure}   

We perform a simulation driven by the same amplitude of external forcing as in section \ref{sec:micro_model}\ref{subsec:microscopic_result} but with $T_s/T_d=5$. The period of convection and wave is still set as 1 day, and the difference of the lower and upper thresholds is still set as $\Delta \Theta = 1$ K. The initial convective phase is uniformly distributed within the shallow and deep stages. The number of convective members at the deep stage is $1/5$ of the shallow stage. This setting makes $\widetilde{\Theta} \approx 0$ at $t=0$ day. Figure \ref{fig:synchro_asymmetry_run=4}a shows that synchronization still occurs but is much slower than the symmetric case reported in section \ref{sec:micro_model}\ref{subsec:microscopic_result}. To explain why the synchronization is suppressed by asymmetry, we consider an extreme case where $T_s \gg T_d$, as illustrated in Fig. \ref{fig:synchro_asymmetry_run=4}b. In this hypothetical case, the accumulated wave forcing at a deep stage is negligible because $T_d$ is very short; the accumulated wave forcing at a shallow stage is always close to zero, irrespective of the phase relation between the wave and convection. As a result, a wave can hardly adjust the convective phase.

Because the asymmetric case is much harder to solve analytically, we leave the question of whether $\widetilde{\Theta}$ still obeys a harmonic oscillator for future work.

\section{Noise}

\begin{figure}[h]
\centerline{\includegraphics[width=1.4\linewidth]{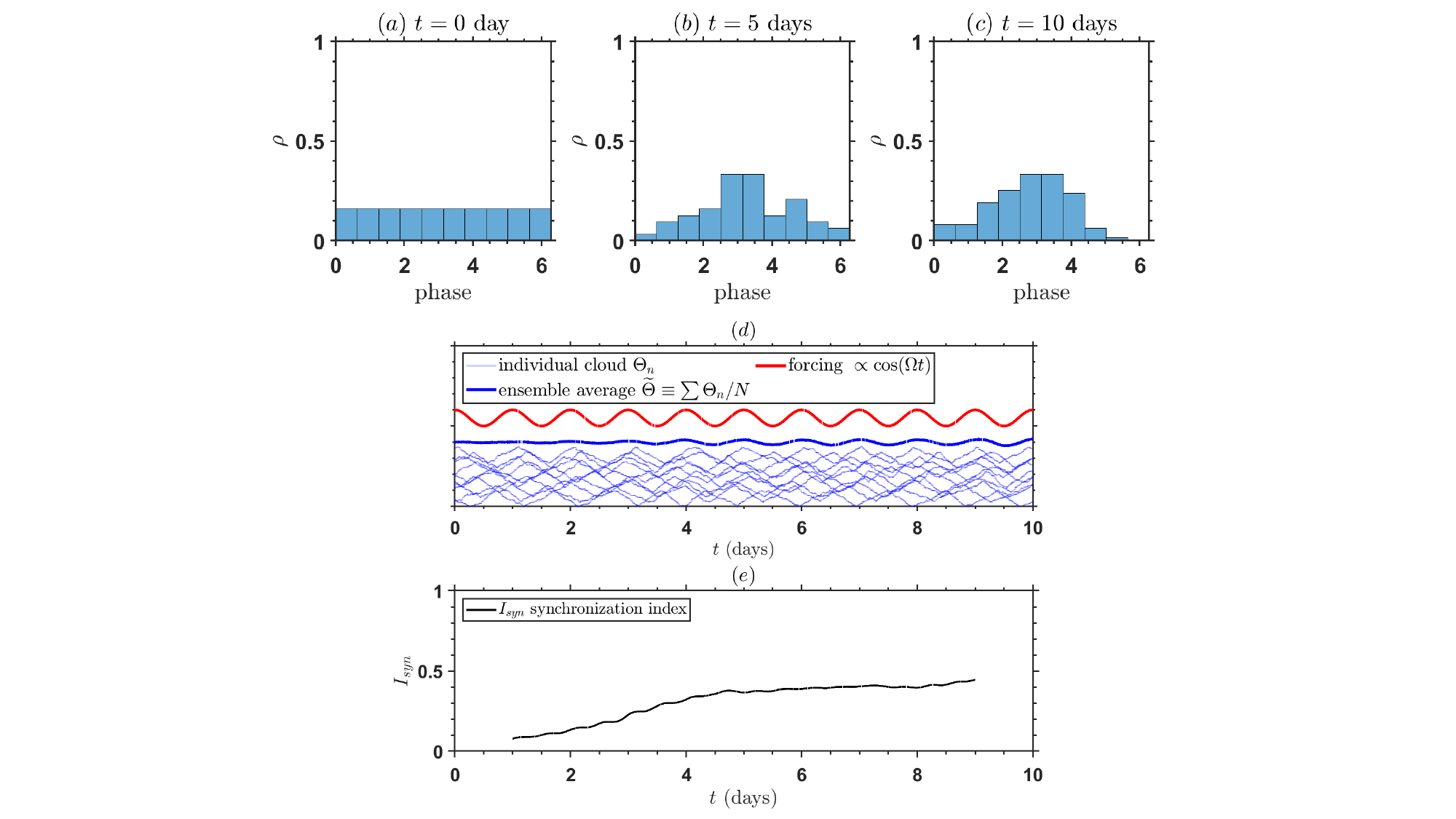}}
\caption{The same as Fig. \ref{fig:synchro_show_hist}, but for the microscopic model simulation with noise. }   \label{fig:synchro_show_hist_noise}
\end{figure}

We perform the same simulation in section \ref{sec:micro_model}\ref{subsec:microscopic_result}, but adding a noise term $\xi$ to the $\Theta_n$ equation:
\begin{equation}
    \frac{d\Theta_n}{dt} = Q_{s,n} + Q_{d,n} + Q_{rad} + Q_{surf} - \widetilde{W} \left\langle \frac{d\theta_{e,0}}{dz} \right\rangle + \xi.
\end{equation}
Here, $\xi$ is set to obey a uniform distribution sampled between $\pm 2\times10^{-4}$ K s$^{-1}$ and is added at every time step (60 seconds). Comparing the simulation result with noise (Fig. \ref{fig:synchro_show_hist_noise}) to that without noise (Fig. \ref{fig:synchro_show_hist}), we see the noise significantly suppresses synchronization and the growth of $\widetilde{\Theta}$. From a microscopic view, the noise suppresses synchronization by disturbing the phase adjustment. From a macroscopic view, the noise deviates the convective frequency from the wave frequency, which breaks the resonance condition. 

\appendix[D]

\appendixtitle{The convective phase distribution function $\rho$}

In this appendix, we quantitatively study the synchronization under a periodic wave forcing. The goal is to analytically solve the convective phase distribution function $\rho$ and the ensemble-averaged equivalent potential temperature $\widetilde{\Theta}$. We start by deriving the phase evolution equation of an individual convective cell.

\subsection{The phase evolution equation}\label{subsec:theta_n}


The time derivative of the convective phase $\varphi_n$ is calculated with its definition [Eq. (\ref{eq:phase_definition})]:
\begin{equation}  
\label{eq:phin_with_wave}
    \frac{d\varphi_n}{dt} = 
    \begin{cases}
        \frac{\pi}{\Delta \Theta} \frac{d\Theta_n}{dt}
        = \omega + \frac{\pi}{\Delta \Theta} \widetilde{W} \left\langle \frac{d\theta_{e,0}}{dz} \right\rangle, \quad 0<\varphi_n<\pi, \\
        - \frac{\pi}{\Delta \Theta} \frac{d\Theta_n}{dt}
        = \omega - \frac{\pi}{\Delta \Theta}\widetilde{W} \left\langle \frac{d\theta_{e,0}}{dz} \right\rangle,\quad \pi<\varphi_n<2\pi,
    \end{cases}
\end{equation}
where we have substituted in the $\Theta_n$ evolution equation (\ref{eq:Theta_simplified}). Here, $\omega = 2\pi/T$ [Eq. (\ref{eq:omega_expression})] is the constant intrinsic angular frequency of convection. The piecewise nature of the wave-induced phase drift in Eq. (\ref{eq:phin_with_wave}) at the shallow and deep stages makes it hard to obtain an accurate solution. To move forward, we simplify the wave-induced phase drift by only calculating its bulk effect at the shallow and deep stages, i.e., a stage-averaged approach.\footnote{The stage-averaged approach was inspired by the derivation of Stokes drift in the textbook of \citet{mcwilliams2006book}.} Assuming the wave is a small perturbation to the convective lifecycle, we take $T/2$ as the integral time over each stage. This assumption neglects the wave's influence on the integral time, which is small for a weakly forced case and might be considered a next-order correction. Equation (\ref{eq:phin_with_wave}) is approximated as:
\begin{equation}
\label{eq:dphi_dt_withwave}
    \frac{d\varphi_n}{dt} \approx \omega
    + \frac{\pi}{\Delta \Theta} \left\langle \frac{d\theta_{e,0}}{dz} \right\rangle \left( \frac{1}{T}\int_{t_n}^{t_n+T/2} \widetilde{W} dt - \frac{1}{T} \int_{t_n+T/2}^{t_n+T} \widetilde{W} dt
    \right).
\end{equation}
Here, $t_n$ is the deep-to-shallow transition time of the studied convective lifecycle.

The wave vertical velocity $\widetilde{W}$ is defined as:
\begin{equation}  \label{eq:wave_forcing_app}
\begin{split}
    \widetilde{W} &= \widetilde{W}_0 \cos \left( \varphi_w + \pi \right) \\
    &= \widetilde{W}_0 \cos \left( \Omega t + \check{\varphi}_w + \pi \right),
\end{split}    
\end{equation}
where $\widetilde{W}_0$ is a constant wave amplitude, and $\Omega$ is the wave angular frequency. Note that $\Omega$ does not necessarily equal the intrinsic angular frequency of convection, $\omega$. The wave phase is $\varphi_w$, a function of time. The initial wave phase is $\check{\varphi}_w$. The $\varphi_w$ is linked to $\check{\varphi}_w$ via:
\begin{equation}  \label{eq:wave_initial_phase}
    \varphi_w = \check{\varphi}_w + \Omega t.
\end{equation}

Substituting Eq. (\ref{eq:wave_forcing_app}) into Eq. (\ref{eq:dphi_dt_withwave}), we obtain:
\begin{equation}
 \label{eq:phase_drift}
\begin{split}
    \frac{d\varphi_n}{dt}
    &= \omega + \frac{\pi}{\Delta \Theta} \widetilde{W}_0 \left\langle \frac{d\theta_{e,0}}{dz} \right\rangle \left[ \frac{1}{T}\int_{t_n}^{t_n+T/2} \cos(\check{\varphi}_w+\Omega t)dt
    - \frac{1}{T}\int_{t_n+T/2}^{t_n+T} \cos(\check{\varphi}_w+\Omega t)dt \right] \\
    &= \omega - \frac{\pi}{\Omega T}\frac{\widetilde{W}_0}{\Delta \Theta} \left\langle \frac{d\theta_{e,0}}{dz} \right\rangle \{ \sin \left[ \check{\varphi}_w + (\Omega - \omega) t_n + \omega t_n   \right] \\
&\,\,\quad \quad \quad \quad \quad \,\,\,\,\,\,  \quad    -2 \sin \left[ \check{\varphi}_w + (\Omega - \omega) t_n + \omega t_n + \pi \Omega/\omega \right] \} \\   &\,\,\,\,\,\,\,\,\quad \,\,\,\,\,\, \quad \quad \quad \quad \quad    + \sin \left[ \check{\varphi}_w + (\Omega - \omega) t_n + \omega t_n + 2 \pi \Omega/\omega \right] \} \\
&\approx \omega -\frac{2\omega}{\Omega}\frac{\widetilde{W}_0}{\Delta \Theta} \left\langle \frac{d\theta_{e,0}}{dz} \right\rangle \sin \left[ \check{\varphi}_w - \check{\varphi}_n + (\Omega - \omega) t  \right], \\
\end{split}
\end{equation}
Here, we have used $\omega \equiv 2\pi / T$ and have defined the initial convective phase as: $\check{\varphi}_n \equiv - \omega t_n$, which is related to $\varphi_n$ via:
\begin{equation}  \label{eq:convection_initial_phase}
     \varphi_n=\check{\varphi}_n + \omega t.
\end{equation}
The assumption of weak deviation of wave frequency from the convective frequency is used in deriving the last line of Eq. (\ref{eq:phase_drift}):
\begin{equation}
    |\Omega - \omega | \ll \omega.
\end{equation}
This leads to two simplifications. First, the weak frequency deviation assumption allows us to combine the three sinusoidal terms into one:
\begin{equation}
\begin{split}
    &\sin \left[ \check{\varphi}_w + (\Omega - \omega) t_n + \omega t_n  \right]  \\ 
    -2 &\sin \left[ \check{\varphi}_w + (\Omega - \omega) t_n + \omega t_n  + \pi \Omega/\omega \right] \\
    + &\sin \left[ \check{\varphi}_w + (\Omega - \omega) t_n + \omega t_n + 2 \pi \Omega/\omega \right]
    \approx 4 \sin \left[ \check{\varphi}_w - \check{\varphi}_n + (\Omega - \omega) t_n \right].
\end{split}
\end{equation}
Second, we notice that the $(\Omega - \omega) t_n$ term is the only time-dependent part of the phase tendency. When $|\Omega - \omega | \ll \omega$, the phase tendency is a slow-varying function. This permits a continuous approximation that replaces $t_n$ with $t$:
\begin{equation}
    4 \sin \left[ \check{\varphi}_w - \check{\varphi}_n + (\Omega - \omega) t_n \right]
    \approx 
    4 \sin \left[ \check{\varphi}_w - \check{\varphi}_n + (\Omega - \omega) t \right].
\end{equation}
Applying Eqs. (\ref{eq:wave_initial_phase}) and (\ref{eq:convection_initial_phase}) to the last line of Eq. (\ref{eq:phase_drift}), we see that in the case of $|\Omega - \omega | \ll \omega$, the phase evolution equation approximately obeys:
\begin{equation}  \label{eq:theta_n_equation}
    \frac{d\varphi_n}{dt}
    \approx 
    \omega - K\sin \left( \varphi_n - \varphi_w \right),
\end{equation}
where $K$ is:
\begin{equation}
    K = - \frac{\omega}{\Omega} \frac{2 \widetilde{W}_0}{\Delta \Theta} \left\langle \frac{d\theta_{e,0}}{dz} \right\rangle.
\end{equation}
Note that $K$ is proportional to the only non-dimensional parameter of the system, $B$ [defined in Eq. (\ref{eq:B})], which denotes the relative strength of wave forcing to the convective lifecycle evolution. Till now, two assumptions have been used: weak forcing and weak frequency deviation.

\subsection{The phase distribution function $\rho$}\label{subsec:rho}

When the member size of the cloud ensemble is sufficiently large, we can define a phase distribution function $\rho(\varphi,t)$, which describes the probability density of convection at a given phase $\varphi$. Because the number of convective cells in the cloud ensemble is assumed conservative, $\rho$ obeys a conservation equation:
\begin{equation}  \label{eq:rho_general}
    \frac{\partial \rho}{\partial t} + \frac{\partial}{\partial \varphi} \left( \rho\frac{d\varphi}{dt} \right) = 0,
\end{equation}
which is essentially a Liouville equation that describes the evolution of a distribution function in the phase space \cite[e.g.,][]{yano2017convective}. Substituting the phase drift equation (\ref{eq:theta_n_equation}) into Eq. (\ref{eq:rho_general}), we get a closed $\rho$ equation:
\begin{equation}  \label{eq:rho_expanded}
\begin{split}
    \frac{\partial \rho}{\partial t}
    &= - \frac{\partial}{\partial \varphi} \left[ \rho \omega - \rho K \sin \left( \varphi - \varphi_w \right) \right]\\
    &= - \omega \frac{\partial \rho}{\partial \varphi}
    + K \sin (\varphi - \varphi_w) \frac{\partial \rho}{\partial \varphi}
    + \rho K \cos \left( \varphi - \varphi_w \right).
\end{split}    
\end{equation}
Assuming the synchronization status is weak (the shape of $\rho$ is relatively uniform with $\varphi$) and the synchronization rate is slow (the wave forcing is weak):
\begin{equation}
    \left| \rho - \frac{1}{2\pi} \right| \ll \frac{1}{2\pi},
    \quad 
    K \ll \omega,
\end{equation}
the second term on the right-hand side of Eq. (\ref{eq:rho_expanded}) can be neglected, so the $\rho$ equation is approximated as:
\begin{equation}  \label{eq:linearized_rho_equation}
    \frac{\partial \rho}{\partial t} + \omega \frac{\partial \rho}{\partial \varphi} 
    \approx \frac{K}{2\pi} \cos \left( \varphi - \varphi_w \right)
    = \frac{K}{2\pi} \cos \left( \varphi - \check{\varphi}_w - \Omega t \right).
\end{equation}
Without loss of generality, we set $\check{\varphi}_w=-\pi$, which makes the wave forcing [Eq. (\ref{eq:wave_forcing_app})] appear as $\widetilde{W}=\widetilde{W}_0 \cos (\Omega t)$, the same as the demonstrative simulation in section \ref{sec:micro_model}\ref{subsec:microscopic_result}. Equation (\ref{eq:linearized_rho_equation}) can be solved by first generalizing it to a complex form:
\begin{equation}  
    \frac{\partial \rho}{\partial t} + \omega \frac{\partial \rho}{\partial \varphi} 
    = - \frac{K}{2\pi} \exp\left[ i (\varphi - \Omega t ) \right],
\end{equation}
and then substituting in a trial solution:
\begin{equation}  \label{eq:rho_trial_solution}
    \rho = \frac{1}{2\pi} + \hat{\rho}(t) \exp\left[ i (\varphi - \omega t) \right].
\end{equation}
Here, $\hat{\rho}(t)$ is the time-dependent amplitude of the histogram fluctuation, whose governing equation is:
\begin{equation}
    \frac{d\hat{\rho}}{dt} = - \frac{K}{2\pi} \exp \left[ i (\omega - \Omega) t \right].
\end{equation}
Using an initial condition of $\hat{\rho}|_{t=0}=0$, the solution of $\hat{\rho}$ is:
\begin{equation} \label{eq:rho_hat_solution}
    \hat{\rho}(t) = \begin{cases}
        - \frac{K}{2\pi} t,\quad \omega = \Omega, \\
        - \frac{K}{2\pi i}\frac{1}{\omega-\Omega} \left\{ \exp \left[ i (\omega - \Omega) t \right] - 1 \right\}, \quad \omega \neq \Omega.
    \end{cases}
\end{equation}
Substituting Eq. (\ref{eq:rho_hat_solution}) into Eq. (\ref{eq:rho_trial_solution}) and taking the real part, we get the solution of $\rho$:
\begin{equation}  \label{eq:rho_solution}
    \rho = \frac{1}{2\pi} + 
    \begin{cases}
        - \frac{K}{2\pi} t \cos(\varphi - \Omega t),\quad \omega = \Omega, \\
        - \frac{K}{2\pi}\frac{1}{\omega - \Omega} \left[ \sin \left( \varphi - \Omega t \right) - \sin \left( \varphi - \omega t \right) \right], \quad \omega \neq \Omega.
    \end{cases}
\end{equation}
Note that no matter whether $\omega$ equals $\Omega$, $\rho$ is always a sinusoidal function of $\varphi$. We will show that this is an important property to link $\rho$ to the ensemble-averaged quantity $\widetilde{\Theta}$.

\subsection{Representing the ensemble average with the phase distribution}\label{subsec:average_rho}

The last step is to represent the ensemble-averaged quantity $\widetilde{\Theta}$ with the phase distribution function $\rho$. We let $f_{\Theta}(\varphi)$ be the shape function of $\Theta_n$ without considering the wave perturbation, which takes a tent shape:
\begin{equation} \label{eq:shape_fun_Theta}
\begin{split}
    f_{\Theta} (\varphi)
    &= \Delta \Theta \left( \frac{1}{2} - \frac{|\varphi-\pi|}{\pi} \right).
\end{split}    
\end{equation}
Here, $\mathcal{H}$ is the Heaviside function. As a convention, $\varphi=0$ and $\varphi=\pi$ are defined as the deep-to-shallow and shallow-to-deep transition phases, respectively. Because the solution of $\rho$ is always a sinusoidal function of $\varphi$ [Eq. (\ref{eq:rho_solution})], we use a few lines of derivation to express $\widetilde{\Theta}$ as:
\begin{equation}  \label{eq:Theta_tilde_expression}
\begin{split}
    \widetilde{\Theta}
    &= \int_{0}^{2\pi} f_{\Theta}(\varphi) \rho(\varphi,t) d\varphi \\
    &= - \frac{2 \Delta \Theta}{\pi} \left[ \rho (\varphi=0) - \rho (\varphi=\pi) \right].    
\end{split}    
\end{equation}
Substituting Eq. (\ref{eq:rho_solution}) into Eq. (\ref{eq:Theta_tilde_expression}), we get the solution of $\widetilde{\Theta}$:
\begin{equation}  \label{eq:analytical_A}
    \widetilde{\Theta}
    =  - \frac{8}{\pi^2} \frac{\widetilde{W}_0}{2} \left\langle \frac{d\theta_{e,0}}{dz} \right\rangle t \cos(\Omega t),\quad \omega = \Omega,
\end{equation}
\begin{equation}  \label{eq:analytical_B}
    \widetilde{\Theta}
    = - \frac{8}{\pi^2}  \Omega \widetilde{W}_0 \left\langle \frac{d\theta_{e,0}}{dz} \right\rangle \frac{1}{\omega - \Omega} \frac{\omega}{\Omega^2} \cos \left( \frac{\omega+\Omega}{2} t \right) \sin \left( \frac{\omega-\Omega}{2} t \right),\quad \omega \neq \Omega.
\end{equation}

\subsection{Ensemble-averaged equation}

We notice that the analytical solution of the $\omega =\Omega$ case is the solution to the oscillator equation:
\begin{equation}  \label{eq:oscillator_appendix}
    \frac{d^2 \widetilde{\Theta}}{dt^2} + \omega^2 \widetilde{\Theta} = - \frac{8}{\pi^2} \frac{d}{dt}\left( \widetilde{W} \left\langle \frac{d\theta_{e,0}}{dz} \right\rangle \right).
\end{equation}
The $8/\pi^2$ factor is associated with the tent function shape of $\Theta_n$, which is close to unity. The $\omega \neq \Omega$ case solution is not exactly the solution of Eq. (\ref{eq:oscillator_appendix}), but we can show it is an approximation. Under the $|\omega-\Omega| \ll \omega$ assumption, which is used in deriving the phase drift equation [Eq. (\ref{eq:phase_drift})], the $\Omega^2/\omega$ factor can be crudely approximated as:
\begin{equation}  \label{eq:Omega_omega_approx}
    \frac{\Omega^2}{\omega}
    \approx \frac{\omega + \Omega}{2}.
\end{equation}
Substituting Eq. (\ref{eq:Omega_omega_approx}) into Eq. (\ref{eq:analytical_B}). We obtain an approximate solution of $\widetilde{\Theta}$ in the $\omega \neq \Omega$ case that obeys the oscillator equation:
\begin{equation}  \label{eq:oscillator_solution_neq}
    \widetilde{\Theta}
    \approx - \frac{8}{\pi^2}  \Omega \widetilde{W}_0 \left\langle \frac{d\theta_{e,0}}{dz} \right\rangle  \frac{2}{\omega^2 - \Omega^2} \cos \left( \frac{\omega+\Omega}{2} t \right) \sin \left( \frac{\omega-\Omega}{2} t \right),\quad \omega \neq \Omega.
\end{equation}
The analytical solution in the $\omega=\Omega$ case [Eq. (\ref{eq:analytical_A})] and $\omega \neq \Omega$ case [Eq. (\ref{eq:oscillator_solution_neq})] are benchmarked with numerical simulations of the microscopic model, as shown in Fig. \ref{fig:analytical_solution_test}.



\bibliographystyle{ametsocV6}
\bibliography{amspaperV6}

\end{document}